\documentclass[lettersize,journal]{IEEEtran}
\usepackage{amsmath,amsfonts}
\usepackage{algorithmic}
\usepackage{algorithm}
\usepackage{array}
\usepackage[caption=false,font=normalsize,labelfont=sf,textfont=sf]{subfig}
\usepackage{textcomp}
\usepackage{stfloats}
\usepackage{url}
\usepackage{verbatim}
\usepackage{graphicx}
\usepackage{cite}
\usepackage{makecell}
\usepackage{mathtools}

\usepackage[export]{adjustbox}
\begin{document}

\title{FMCW Radar Interference Mitigation based\\ on the Fractional Fourier Transform}

\author{Christian Oswald, Franz Pernkopf
\thanks{Christian Oswald and Franz Pernkopf are with the Signal Processing and Speech
Communication Laboratory, Graz University of Technology, 8010 Graz, Austria (e-mail: christian.oswald@tugraz.at, pernkopf@tugraz.at)}
\thanks{Research funded by the Austrian Research Promotion Agency (FFG), Infineon Technologies Austria AG and Graz University of Technology under the REPAIR project (40352729)}
\thanks{© 2026 IEEE. Personal use of this material is permitted. Permission from IEEE must be obtained for all other uses, in any current or future media, including reprinting/republishing this material for advertising or promotional purposes, creating new collective works, for resale or redistribution to servers or lists, or reuse of any copyrighted component of this work in other works.}
\thanks{This work has been published in IEEE Transactions on Radar Systems, vol. 4, pp. 549 - 563, 2026. doi: 10.1109/TRS.2026.3666094.}}

\maketitle
\thispagestyle{empty}

\begin{abstract}
In this paper, we propose a novel method for frequency modulated continuous wave (FMCW) radar mutual interference mitigation (IM) based on the discrete fractional Fourier transform (DFrFT). Interference chirps are detected and mitigated by compression and zeroing in the fractional domain. We provide an efficient implementation that can deal with multiple interferers, where we perform consecutive DFrFTs utilizing its angle-additivity property. For that purpose, we generalize and reduce the computational complexity of the multi-angle centered discrete fractional Fourier transform \cite{vargas2005multiangle}. Our algorithm is designed to be simple and fast such that it can be implemented in hardware. 
We evaluate our algorithm on a synthetic I/Q-modulated dataset and outperform reference methods in terms of the mean squared error, signal-to-interference-plus-noise ratio, error vector magnitude, true positive rate, false alarm rate and F1-score.
\end{abstract}


\section{Introduction} \label{sec:intro}
\IEEEPARstart{F}{MCW} radar has established itself as an indispensable component of advanced driver assistance systems and autonomous vehicles due to its low price, long range, velocity measuring abilities, independence from weather and lighting conditions, among other advantages. However, as the number of radar systems deployed increases, radar sensors interfering with one another becomes a pressing issue. If ignored, mutual interference may drastically deteriorate object detection performance, as it can appear as noise or even ghost objects in the radar sensor's output. 

\begin{figure}[!t]
\includegraphics[width=\columnwidth]{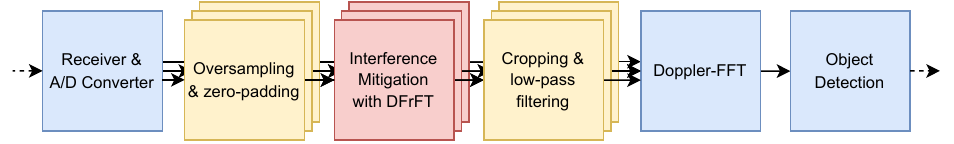}
\caption{FMCW radar signal processing chain with our proposed IM algorithm in red, which replaces the range-FFT. It processes one fast-time sequence at a time and outputs a range-spectrum. The yellow boxes represent our padding scheme proposed in Sec. \ref{sec:padding}, which is optional and can be bypassed. \label{fig:proc_chain}}
\end{figure}

FMCW radar mutual interference is a well-studied problem and has already been tackled with a multitude of countermeasures.
Methods like frequency hopping \cite{bechter2016bats} try to avoid interference {altogether} by switching the sensor's transmit parameters as soon as interference has been detected. Other methods, including our proposed algorithm, mitigate interferences by removing them from the sensor's output as a post hoc process. Such algorithms can be categorized by their placement in the FMCW radar signal processing chain visible in Fig. \ref{fig:proc_chain}, i.e., whether they are applied to fast-time/slow-time sequences, range-spectra, range-Doppler maps, range-Doppler-angle maps or any variations thereof. Zeroing \cite{fischer2016untersuchungen} is a simple and popular IM technique, where interfered samples in a fast-time sequence are detected and zeroed. An iterative method with adaptive thresholding \cite{marvasti2012sparse} can be used on a previously zeroed signal to reconstruct its sparse range-spectrum. Variational signal separation of a fast-time sequence based on sparse Bayesian learning is proposed in \cite{toth2024variational}. The short-time Fourier transform (STFT) of a fast-time sequence is processed in{\cite{muja2022interference, wang2021cfar, ristea2020fully}; {in} \cite{muja2022interference} they compute order-statistics while \cite{wang2021cfar} uses a constant false alarm rate (CFAR) detector on the STFT followed by zeroing, and \cite{ristea2020fully} transforms the STFT to a range-spectrum using a convolutional neural network (CNN).} 
Ramp filtering \cite{wagner2018threshold} applies a non-linear filter across a set of range-spectra. An adaptive noise canceler processing range-spectra is proposed in \cite{jin2019automotive}. 

References \cite{rock2021resource, fuchs2021complex} use fully convolutional NNs on range-Doppler maps with real-valued and complex-valued activations, respectively. An extension of the latter is described in \cite{fuchs2024multiantenna}, which jointly processes all receive antennae. In \cite{oswald2023angle}, a CNN with three-dimensional convolutions operating on range-Doppler-angle maps is proposed, which requires fewer parameters and generalizes better than \cite{fuchs2024multiantenna}. An improvement of \cite{oswald2023angle} is presented in \cite{oswald2023end}, which introduces separable convolutions and propagates gradients through the object detector while training the NN. FMCW radar is a safety critical technology, where faulty behavior must be avoided under all circumstances. In this work, we therefore prefer model-based over data-driven algorithms such as NNs, since their robustness has not yet been proven in this application. 
\IEEEpubidadjcol

The fractional Fourier transform (FrFT) and closely related techniques such as the chirplet transform and matched filtering have already been used for linearly frequency modulated (LFM) chirp {interference} mitigation. Global navigation satellite system (GNSS) chirp IM by estimating the interference chirp's parameters with the FrFT is proposed in \cite{sun2024novel}; {these} are then used to generate a local LFM signal to dechirp and notch-filter the interfered GNSS signal. In \cite{nafchi2021mitigating}, the time-varying Doppler shift in wireless communications of moving objects is mitigated using the DFrFT. Chirp IM in high-frequency surface wave radar is discussed in \cite{wang2020mitigation}, where they null interferences in the fractional domain and then reconstruct the signal with an autoregressive (AR) model. Furthermore, they propose a recursive least-squares adaptive (RLS) filter in the fractional domain to treat interfered signals. In \cite{zhou2019fractional}, interferences are suppressed by performing a singular value decomposition (SVD) of the Hankel matrix derived from the interfered signal's optimal fractional representation. The parameters of LFM interferences {with time-variant angles of arrival} are estimated with the FrFT in \cite{cui2014wideband}, which are then suppressed using subspace projection techniques. FMCW mutual IM using a reduced chirplet transform {and orthogonal matching pursuit (OMP)} is proposed in \cite{correas2019sparse}. FMCW interference in OFDM radars is filtered using coarse-to-fine dechirping in \cite{maeda2023fmcw}. FMCW mutual interference is compressed and removed using an estimated matched filter in \cite{rameez2022interference}. Recently, \cite{chen2024interference} have proposed a method for FMCW mutual IM using the fast approximate FrFT \cite{ozaktas1996digital}, where they use the golden-section search (GSS) to determine the interference's chirp rate. We discuss the conceptual similarities and differences between \cite{chen2024interference} and our approach in Sec. \ref{sec:relation_chen_et_al}, and compare their performance in Sec. \ref{sec:experiments}. {A summary of methods which pulse-compress LFM chirp interferences is shown in Tab. \ref{tab:related}.

\begin{table*}
    \centering
    {\caption{Comparison of pulse-compressing methods for LFM chirp IM}
    \begin{tabular}{|c||c|c|c|c|}
        \hline
        \textbf{Method} & \makecell{\textbf{used technique for} \\ \textbf{pulse-compression}} & \makecell{\textbf{Search for LFM} \\ \textbf{Chirp Parameters}} & \makecell{\textbf{Mitigation} \& \textbf{Interpolation}} & \textbf{Inverse Transform after IM} \\
       \hline \hline
         \cite{sun2024novel}            & \makecell{approx. FrFT \cite{ozaktas1996digital} \& dechirping} & iterative & notch filter & rechirping \\
        \hline
         \cite{wang2020mitigation}      & approx. FrFT \cite{ozaktas1996digital} & chirp rate known & \makecell{zeroing \& AR model or adaptive RLS} & inverse approx. FrFT \\
        \hline
         \cite{zhou2019fractional}      & approx. FrFT \cite{ozaktas1996digital} & chirp rate known  &  \makecell{projection \& SVD of Hankel matrix} & inverse approx. FrFT \\
         \hline
         \cite{cui2014wideband}        & \makecell{approx. FrFT \cite{ozaktas1996digital}} & not discussed & \makecell{subspace projection} & not needed \\
         \hline
         \cite{correas2019sparse}       & \makecell{chirplet transform}  & \makecell{OMP (iterative)} & subtract reconstructed interference & not needed \\
         \hline
         \cite{maeda2023fmcw}           & dechirping & \makecell{coarse-to-fine (iterative)} & notch filter & rechirping \\
         \hline
         \cite{rameez2022interference}  & \makecell{matched filtering \\    } & duration-based estimation & \makecell{zeroing \& AR model} & matched filter \\
         \hline
         \cite{chen2024interference}    & approx. FrFT \cite{ozaktas1996digital} & GSS (iterative) & \makecell{zeroing \& average amplitude insertion} & inverse approx. FrFT \\
         \hline
         ours                & padded exact EMDFrFT & \makecell{maximum of EMDFrFT} & zeroing & absorbed into EMDFrFT \\
         \hline
    \end{tabular}}
    \label{tab:related}
\end{table*} 
}

In this paper, we propose a novel mitigation algorithm for FMCW radar mutual interference. More concretely,
\begin{enumerate}
    \item {we} generalize and reduce the computational complexity of the multi-angle centered discrete fractional Fourier transform (MDFrFT) \cite{vargas2005multiangle} resulting in our efficient MDFrFT (EMDFrFT). 
    \item We use the EMDFrFT as the core element for a new and simple IM algorithm that can deal with multiple interferences and integrates into the FMCW radar signal processing chain.
    \item We consider the imperfections of current implementations of eigendecomposition-based discrete FrFTs (DFrFTs) and propose a simple signal padding scheme that greatly increases their chirp compression capabilities. 
    \item We conduct experiments comparing our algorithm to reference methods and show performance improvements across all metrics evaluated.
\end{enumerate}
{While the core idea of using the FrFT for LFM chirp signal processing is not new, our proposed method differs from existing literature by providing a concrete algorithm which is tailored to the requirements of automotive radar. More specifically,

\begin{itemize}
    \item we use an exact DFrFT implementation and our new padding scheme (see Sec. \ref{sec:padding}) to increase reliability for safety critical applications. Related work, as listed in Tab. \ref{tab:related}, uses the fast approximate FrFT without padding, which leads to inaccuracies as described in Sec. \ref{sec:relation_chen_et_al}.
    \item We optimize our method for real-time and resource-constrained applications. More concretely, we introduce the EMDFrFT, which reduces computational complexity and allows us to detect and compress interferences in a parallelized manner using the same computations; this replaces iterative search strategies and distinct interference detectors used in the literature. Furthermore, we absorb the computation of inverse DFrFTs and range-spectra into the EMDFrFT (see Sec. \ref{sec:emdfrft_int_mit}) -- an optimization which is exclusive to our EMDFrFT.  
\end{itemize}
}

{This paper is structured as follows: First, we give a brief introduction to the signal model and the DFrFT in Sec. \ref{sec:background}. In Sec. \ref{sec:mit} we develop a high-level algorithm, which we optimize and analyze in Sec. \ref{sec:implementation}. We relate our algorithm to the landscape of IM methods in Sec. \ref{sec:properties}, before describing and conducting experiments in Sec. \ref{sec:experimental_setup} and Sec. \ref{sec:experiments} respectively.} Finally, we conclude and describe potential future work in Sec. \ref{sec:conclusion}. 

Throughout this paper, we use bold capital letters to denote matrices and bold lower case letters for vectors and sets. $\boldsymbol{A}[n,m]$ references the element in row $n$ and column $m$ of matrix $\boldsymbol{A}$. $\boldsymbol{A}[n]$ indexes the entire $n$\textsuperscript{th} row, while $\boldsymbol{b}[m]$ denotes the $m$\textsuperscript{th} sample of time-discrete signal $\boldsymbol{b}$.

\section{Background} \label{sec:background}
\subsection{FMCW Radar}
An FMCW radar sends out LFM chirps, also called frequency ramps, and receives reflections from objects as time-delayed versions of its transmit signal. I/Q-mixing the transmit with the receive signal and sampling in intervals $T_s$ reveals $N_O$ objects as sinusoids \cite{stove1992linear}
\begin{equation}
    \boldsymbol{s}_O[n] = \sum_{i=1}^{N_O} A_i e^{j(\omega_i nT_s+\phi_i)},
\end{equation}
where $A_i, \phi_i$ are an object's amplitude and initial phase, respectively; an object's range is proportional to its frequency $\omega_i$. The radial velocity of objects can then be determined by evaluating the object signal's change of phase over consecutive chirps, which are also termed fast-time/slow-time sequences. The two-dimensional discrete Fourier transform (DFT) of a fast-time/slow-time sequence is a so-called range-Doppler (RD) map, where objects appear as peaks with coordinates corresponding to their respective ranges and velocities. Azimuth and elevation of objects can be measured by jointly processing multiple receive antennae. In this paper, clutter and all sources of noise are collected in $\boldsymbol{s}_\mathcal{N}$, which is modeled as complex-valued zero-mean additive white Gaussian noise.

\subsection{Mutual Interference in FMCW Radar} \label{sec:sig_model}
When multiple FMCW radar sensors transmit in the same frequency range, mutual interference might occur. More concretely, an interfering radar's frequency course will be visible in the victim radar's output signal while it crosses its receive frequency band. 
After I/Q-demodulation, ideal anti-aliasing filtering with bandwidth $B$ and sampling, an interference chirp $\boldsymbol{s}_I$ is given as

\begin{equation}
{\boldsymbol{s}_I[n]} =
\begin{cases}
A e^{j(-2\pi k\tau n T_s + \pi k n^2 T_s^2+ \phi_0)} & \frac{\tau-B/k}{T_s} < n <\frac{\tau+B/k}{T_s},\\
{0,}&{\text{otherwise{,}}} 
\end{cases} \label{eq:int_model}
\end{equation}
where $A$, $k$, and $\phi_0$ are the interference's amplitude, chirp rate, and initial phase, respectively. $\tau$ denotes the point in time at which the frequency courses of the interferer and victim radar cross. An interference's chirp rate $k$ is calculated as $k = B_I/T_I - B_V/T_V$, where $B_V,B_I,T_V,T_I$ are half of the victim and interferer transmit bandwidth and ramp duration, respectively. The interference is also an LFM chirp, which is suppressed as soon as its instantaneous frequency is greater than $B$. 
Sometimes, the interferer or victim fast-time sequence ends before the interference chirp crosses the entire receiver bandwidth; {this} case is not considered in \eqref{eq:int_model}, but we discuss such interferences and their effect on our proposed method in Sec. \ref{sec:incomplete}. A more detailed description of FMCW mutual interference can be found in \cite{toth2018analytical}. 

\begin{figure*}[!t]
\centering
\includegraphics[width=\textwidth]{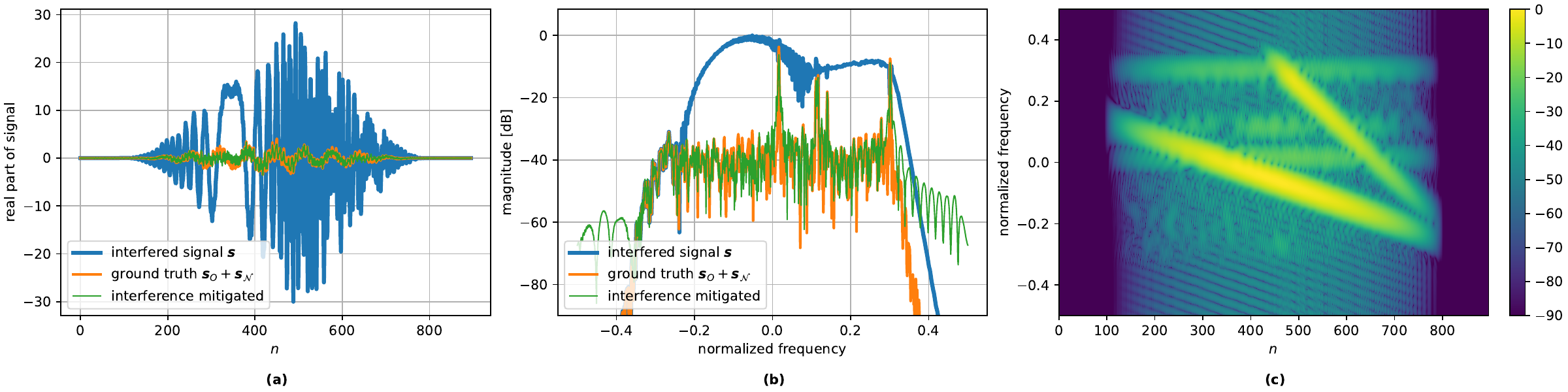}
\caption{Example of an FMCW radar signal with four objects and two interferences. The signal has been padded using the technique introduced in Sec. \ref{sec:padding}. \textbf{(a)} real parts of the time-domain ground truth signal as well as the signals before and after {IM} using our method. \textbf{(b)} corresponding normalized range-spectra. \textbf{(c)} corresponding STFT of interfered signal. The objects are visible as horizontal and the interferences as tilted lines.} \label{fig:sig_ex}
\end{figure*}

In this paper, radar signals $\boldsymbol{s}$ are modeled as a superposition of $N_I$ interferences $\boldsymbol{s}_{I}$, an object signal $\boldsymbol{s}_O$ and noise $\boldsymbol{s}_\mathcal{N}$,  
\begin{equation}
    \boldsymbol{s} = \sum_{m=1}^{N_I}  \boldsymbol{s}_{I_m} +\boldsymbol{s}_O + \boldsymbol{s}_\mathcal{N}. \label{eq:superposition}
\end{equation}
An example for such a signal can be seen in Fig. \ref{fig:sig_ex}.

\subsection{The Fractional Fourier Transform} \label{sec:frft_background}
The fractional Fourier transform (FrFT) is a generalization of the Fourier transform (FT), as it interpolates between a time-domain signal and its spectrum. 
It is defined as the $a$\textsuperscript{th} power of the FT operator $\mathcal{F}$, $a \in \mathbb{R}$ being the so-called fractional order. For $a = 0$ the FrFT becomes the identity function, for $a = -1$ the inverse FT and for $a = 2$ the parity operator. $\mathcal{F}$ has a periodicity of $4$, as $\mathcal{F}^{a + 4h} =\mathcal{F}^a, \forall h \in \mathbb{Z}$. Intuitively, a forward or inverse FT can be seen as a rotation of a signal's Wigner-Ville distribution \cite{boashash2015time} by $90^\circ$ or $-90^\circ$, respectively. The FrFT extends this notion of rotation to all other angles. The basis functions of the FrFT are LFM chirps with chirp rates parameterized by $a$. We define the fractional angle $\alpha = a \pi/2$ such that we can describe a FrFT $\mathcal{F}^{\frac{2\alpha}{\pi}} \coloneq \mathcal{F_\alpha}$ by its rotation angle of the time-frequency plane. The estimation of LFM chirp rates and center frequencies using the FrFT was shown to be asymptotically unbiased and achieves the Cramer-Rao lower bound \cite{aldimashki2020performance}.
In addition to its reduction to the FT for $\alpha=90^\circ$, the FrFT has two main properties which we will use in the development of our algorithm:
\begin{enumerate}
    \item \textit{Angle-additivity:} $\mathcal{F}_{\alpha_{1}} \circ\mathcal{F}_{\alpha_{2}} = \mathcal{F}_{\alpha_{2}} \circ\mathcal{F}_{\alpha_{1}} = \mathcal{F}_{\alpha_{2} +\alpha_{1}}$, {where $\circ$ indicates function composition.} \label{eq:angle_additivity}
    \item \textit{Unitarity:} $(\mathcal{F}_\alpha)^{-1} = \mathcal{F_{-\alpha}} = (\mathcal{F_{\alpha}})^H$, \label{eq:unitarity} where $H$ indicates Hermitian conjugation. It follows that Parseval's theorem extends from the FT to the FrFT, i.e., 
    \begin{equation}
        \int_{-\infty}^{\infty}|x(t)|^2 dt = \int_{-\infty}^{\infty}|\mathcal{F}_{\alpha} \{x(t)\}(u)|^2 du, \label{eq:parseval}
    \end{equation}
    for any $\alpha \in \mathbb{R}$.
\end{enumerate}
The adaptation of the FrFT to time-discrete signals is called the discrete fractional Fourier transform (DFrFT) $\boldsymbol{W}^{\frac{2\alpha}{\pi}} \coloneq \boldsymbol{W}_{\alpha}$, which is defined as the fractional power of the DFT matrix $\boldsymbol{W}$. However, there exist different implementations of the DFrFT as the eigendecomposition
\begin{equation}
    \boldsymbol{W}^{\frac{2\alpha}{\pi}} = \boldsymbol{V} \boldsymbol{\Lambda}^{\frac{2\alpha}{\pi}} \boldsymbol{V}^T,
\end{equation} 
into the DFT eigenvectors $\boldsymbol{V}$ and eigenvalues $\boldsymbol{\Lambda}$ is not unique. 
Different implementations of the DFrFT provide different advantages and are still subject of current research.
From an application point of view, the most important consideration when choosing a DFrFT implementation is its computational complexity and its properties needed in the application. 
Sampling-based DFrFT implementations such as \cite{ozaktas1996digital, pei2000closed} utilize FFTs and therefore have complexity $\mathcal{O}(N \log N)$ for a signal of length $N$; {however}, they do not have the angle-additivity property, and the method in \cite{ozaktas1996digital} is not unitary. On the other hand, 
implementations such as eigendecomposition-based DFrFTs \cite{candan2000discrete, pei1999discrete, serbes2011discrete, de2017discrete, serbes2011efficient} do retain these properties, but they are computed as a matrix multiplication and therefore have complexity $\mathcal{O}(N^2)$. A survey of existing DFrFT implementations can be found in \cite{su2019analysis, gomez2020fractional}. There also exist algorithms where the eigendecomposition-based multi-angle DFrFT (MDFrFT) can be computed in $\mathcal{O}(N^2 \log N)$ as opposed to a naive implementation with complexity $\mathcal{O}(N^3)$ \cite{vargas2005multiangle, oswald2026onthe}; we review the MDFrFT in Sec. \ref{sec:implementation} as we generalize it for our algorithm. A comparative study of different centered DFrFTs can be found in \cite{peacock2013comparison}.  

\section{Interference Mitigation using the DFrFT}\label{sec:mit}
\begin{figure}[!t]
\includegraphics[width=\columnwidth]{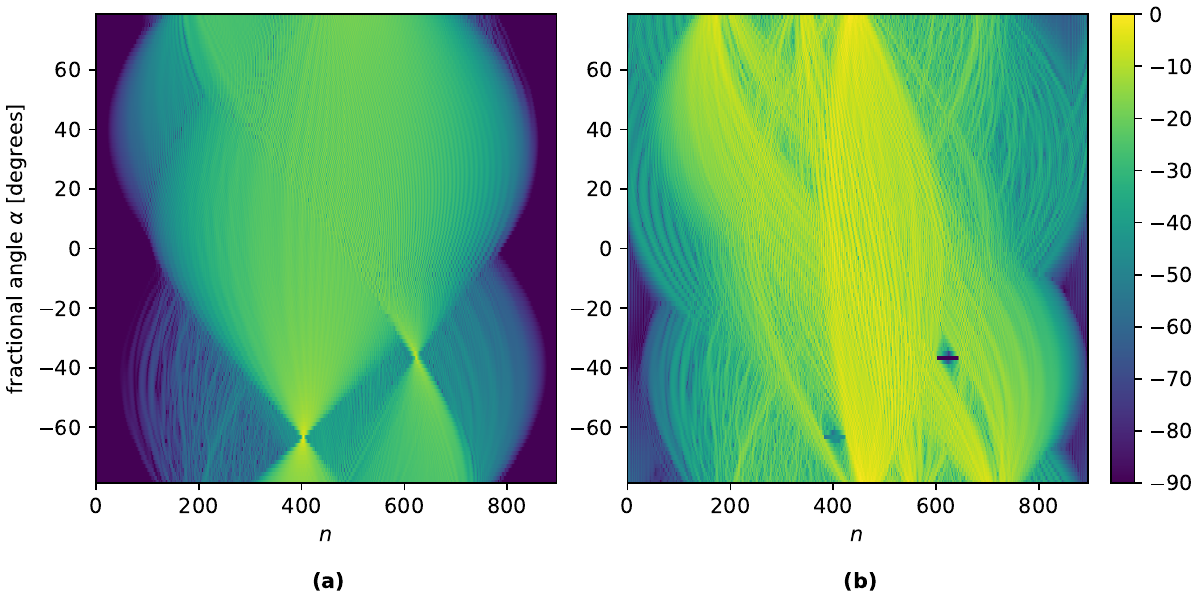}
\caption{DFrFT Magnitudes with angles $\boldsymbol{\alpha}$ of {the} signals in Fig. \ref{fig:sig_ex}. 
All plots are in dB and normalized such that the maximum value is 0. 
\textbf{(a)} interfered signal $\boldsymbol{s}$. The two peaks correspond to the two interferences. \textbf{(b)} interference mitigated signal. {The angular resolution, as defined in Sec. \ref{sec:emdfrft_int_mit}, is $N_{\boldsymbol{\alpha}} = 113$.}}
\label{fig:frfts}
\end{figure}

We now introduce an algorithm which detects and zeroes one interference chirp at a time using the DFrFT. More precisely, our algorithm performs $N_I + 1$ iterations for a radar signal \eqref{eq:superposition} that is corrupted by $N_I$ interference chirps (given that all interferences are detected and zeroed as intended). We first show how our method would compress a pure chirp signal $\boldsymbol{s}_I$ and then extend this approach to radar signals $\boldsymbol{s}$ which additionally contain objects, noise and possibly other interference chirps. 

As described in Sec. \ref{sec:sig_model}, FMCW mutual interference consists of LFM chirps. 
Therefore, a DFrFT with a specific unknown fractional angle $\hat\alpha_I$ will compress a pure interference chirp signal $\boldsymbol{s}_I$ into a pulse with its maximum located at index $\hat n_I$. 
Since the DFrFT is energy preserving \eqref{eq:parseval}, transforming the interference signal with $\boldsymbol{W}_{\hat \alpha_I}$ will result in the highest possible peak $|\boldsymbol{W}_{\hat \alpha_I} \boldsymbol{s}_I|[\hat n_I]$ among all possible values for $\alpha$ and $n$,
\begin{equation}
    \hat\alpha_I, \hat n_I = \underset{\alpha, n}{\text{arg\,max}} (|\boldsymbol{W}_\alpha \boldsymbol{s}_I|[n]). \label{eq:argmax} 
\end{equation}
An example for $\boldsymbol{s}_I$ can be seen in Fig. \ref{fig:padding}a, where $\hat \alpha_I \approx -60^\circ$ and $\hat n_I \approx 400$. 
We perform a grid search for $\hat \alpha_I$ and $\hat n_I$, that is, we search within $N_{\boldsymbol{\alpha}}$ uniformly spaced fractional angles $\boldsymbol{\alpha}$ between $\alpha_{\textrm{max}}$ and $-\alpha_{\textrm{max}}$, where $N_{\boldsymbol{\alpha}}$ and $\alpha_{\textrm{max}}$ are hyperparameters of our method.
A grid search is sufficient, as finding the exact value of $\hat \alpha_I$ is desirable but not necessary for our algorithm; a deviation between the found and the optimal $\hat \alpha_I$ simply corresponds to a weaker compression of the interference. {We discuss and evaluate the influence of $N_{\boldsymbol{\alpha}}$ on our method's performance in Sec. \ref{sec:ang_res} and Sec. \ref{sec:n_angles}, respectively.}
We compute DFrFTs with angles $\boldsymbol{\alpha}$ 
using an efficient and generalized version of the MDFrFT \cite{vargas2005multiangle}, which we introduce in Sec. \ref{sec:MAFrFT}.

In our algorithm, we use \eqref{eq:argmax} to compress and detect an interference chirp $\boldsymbol{s}_I$ within a radar signal $\boldsymbol{s}$. 
For analysis purposes, we split $\boldsymbol{s}$ into the interference chirp we want to compress in a given iteration of our method and a residual signal $\boldsymbol{s}_R$, i.e., $\boldsymbol{s} =\boldsymbol{s}_{I} + \boldsymbol{s}_R$.
Eq. \eqref{eq:argmax} applied to $\boldsymbol{s}$ and $\boldsymbol{s}_R$ returns $\hat\alpha$, $\hat{n}$, $\hat\alpha_R$ and $\hat{n}_R$, respectively; note that in practice, we only have access to $\boldsymbol{s}$. If
\begin{equation}
    |\boldsymbol{W}_{\hat{\alpha}_{I}} \boldsymbol{s}_{I}|[\hat{n}_{I}] > |\boldsymbol{W}_{\hat{\alpha}_R} \boldsymbol{s}_R|[\hat{n}_R], \ \ \ \ \hat\alpha_{I}, \hat\alpha_R \in \boldsymbol{\alpha} \label{eq:main_condition}
\end{equation}
then $\hat{\alpha} \approx \hat{\alpha}_{I}$ and $\hat{n} \approx \hat{n}_{I}$, which means that applying \eqref{eq:argmax} to $\boldsymbol{s}$ essentially returns the same result as when applied to $\boldsymbol{s}_I$.
The superposition of $\boldsymbol{s}_{I}$ and $\boldsymbol{s}_{R}$ might cause deviations of the estimated $\hat{\alpha}, \hat{n}$ from the sought $\hat{\alpha}_{I}, \hat{n}_{I}$; however, these deviations are negligible if $|\boldsymbol{W}_{\hat\alpha_I}\boldsymbol{s}_{I}|[\hat{n}_I] \gg |\boldsymbol{W}_{\hat\alpha_I}\boldsymbol{s}_{R}|[\hat{n}_I]$, which is mostly the case in practice.
An example for a signal where \eqref{eq:main_condition} holds is depicted in Fig. \ref{fig:frfts}a, with $\boldsymbol{s}_I$ being the interference signal from Fig. \ref{fig:padding}a superimposed with objects, noise, and another interference. 

In practice, we do not have access to $\boldsymbol{s}_{I}$ and $\boldsymbol{s}_{R}$ and therefore cannot verify whether \eqref{eq:main_condition} is true for a found peak $(\boldsymbol{W}_{\hat\alpha} \boldsymbol{s})[\hat{n}]$. In other words, we need a different approach to confirm that the global maximum $(\boldsymbol{W}_{\hat\alpha} \boldsymbol{s})[\hat{n}]$ is caused by an interference and not by objects and noise.
Therefore, we use a CFAR-detector: If the global maximum's power exceeds a predefined threshold $\beta$ compared to a reference $\hat{\sigma}^2$, we classify it as interference, that is, we determine that \eqref{eq:main_condition} holds. We choose the average power of $\boldsymbol{s}_R$ as the reference $\hat{\sigma}^2$, which is estimated by the CFAR detector using a window of size $\Phi$ to either side of $(\boldsymbol{W}_{\hat\alpha} \boldsymbol{s})[\hat{n}]$,
\begin{align}
    \hat\sigma^2 &= \frac{1}{2 \Phi}\sum_{n \in \textrm{window}} (|\boldsymbol{W}_{\hat{\alpha}} \boldsymbol{s}|[n])^2 \approx
    \frac{1}{2\Phi}\sum_{n \in \textrm{window}} (|\boldsymbol{W}_{\hat{\alpha}} \boldsymbol{s}_R|[n])^2 \label{eq:cfar_p1}\\
    &\approx\frac{1}{N}\sum_{n=0}^{N-1} (|\boldsymbol{W}_{\hat{\alpha}} \boldsymbol{s}_R|[n])^2 = \frac{1}{N}\sum_{n=0}^{N-1} |\boldsymbol{s}_R[n]|^2. \label{eq:unitary_cfar}
\end{align}
{Note that in \eqref{eq:unitary_cfar} we utilize the DFrFT's unitary property; furthermore,} within the CFAR detector's window $\boldsymbol{W}_{\hat{\alpha}} \boldsymbol{s} \approx \boldsymbol{W}_{\hat{\alpha}} \boldsymbol{s}_R$ because $\boldsymbol{W}_{\hat{\alpha}}\boldsymbol{s}_I$ being sparse, which we use in \eqref{eq:cfar_p1}. We estimate the average power of $\boldsymbol{s}_R$ instead of the noise $\boldsymbol{s}_\mathcal{N}$, which helps us to distinguish between global maxima caused by interferences and objects, as we will explain in Sec. \ref{sec:objs_vs_ints}. The estimate $\hat{\sigma}^2$ contains the energy of objects, noise, and other interferences. Objects increasing $\hat{\sigma}^2$ are not an issue in practice as they only lead to missed detections of interferences that are significantly weaker than these objects. To minimize the influence of other interferences on $\hat\sigma^2$ corrupting some regions of $\boldsymbol{W}_{\hat\alpha}\boldsymbol{s}$ (see, for example, Fig. \ref{fig:frfts}a)
, we use a least-of CFAR (LO-CFAR) detector, where we compute a separate estimate for either side of $(\boldsymbol{W}_{\hat\alpha}\boldsymbol{s})[\hat{n}]$ and then pick the lower one.
We also place $G$ guard cells on either side of the global maximum to deal with the imperfect compression of the interference, which is caused by windowing, the anti-aliasing filter, the DFrFT implementation, interferences not crossing the entire receiver bandwidth (as described in Sec. \ref{sec:incomplete}) and the time-discrete nature of $\boldsymbol{s}$. These guard cells are excluded from the estimation of the residual signal's average power.

If the global maximum has been classified as interference, we can remove it 
by setting $(\boldsymbol{W}_{\hat\alpha}\boldsymbol{s})[\hat{n}]$ and the surrounding guard cells to zero, i.e., we compute $\boldsymbol{d} \odot (\boldsymbol{W}_{\hat\alpha}\boldsymbol{s})$, where $\odot$ is an element-wise multiplication and $\boldsymbol{d}$ a binary mask returned by the CFAR detector. Note that by zeroing we also remove $(\boldsymbol{W}_{\hat\alpha}\boldsymbol{s}_R)[\hat{n}]$ in addition to $(\boldsymbol{W}_{\hat\alpha}\boldsymbol{s}_{I})[\hat{n}]$ as a side effect.

After removing $\boldsymbol{s}_I$, we can retrieve the corresponding time-domain signal by evaluating $\boldsymbol{W}_{-\hat{\alpha}} (\boldsymbol{d} \odot (\boldsymbol{W}_{\hat\alpha}\boldsymbol{s}))$, and then loop the process introduced above to search for more interference chirps within that signal. As described in Sec. \ref{sec:implementation}, we simplify our algorithm by skipping the inverse DFrFT using its angle-additivity property and compute \eqref{eq:argmax} directly on $\boldsymbol{d} \odot (\boldsymbol{W}_{\hat\alpha}\boldsymbol{s})$ in the next iteration of our method. We exit this loop and terminate our algorithm once it does not detect any other interference chirp, i.e., when the energy of the global maximum {compared to $\hat{\sigma}^2$} drops below the CFAR detector's threshold $\beta$; this is the case in Fig. \ref{fig:frfts}b. 
As we repeatedly search for the global maximum in a signal's set of DFrFTs, the algorithm removes interference chirps sorted by their energy, starting with the most energetic. 
The complete algorithm including all optimizations introduced in Sec. \ref{sec:implementation} is summarized in Alg. \ref{alg1}. 

\subsection{Distinguishing Objects from Interferences} \label{sec:objs_vs_ints}
In our algorithm, the only distinction between interference chirps and objects is their chirp rate, with objects having a chirp rate of zero, i.e., being constant frequencies. 
If we were to apply \eqref{eq:argmax} to radar signal that only contains weak interferences or none at all, we would find that for $\alpha_{\textrm{max}} \geq 90^\circ$, $\hat{\alpha} = \pm 90^\circ$ corresponding to the range-spectrum compressing objects into peaks. 
We prevent the false classification of objects as interferences by setting $\alpha_{\textrm{max}} < 90^\circ$.
In other words, we shrink $\boldsymbol{\alpha}$ such that \eqref{eq:main_condition} also holds for $\boldsymbol{s}$ where the energy of $\boldsymbol{s}_{I}$ is smaller than some given fraction of the strongest object's energy. 
From a statistical point of view, interferences with lower chirp rates become increasingly unlikely \cite{hahmann2018evaluation}, which means that we can choose $\alpha_{\textrm{max}}$ slightly smaller than $ 90^\circ$ and still sufficiently compress most interferences. The choice of $\alpha_{\textrm{max}}$ is a trade-off between minimizing false positives versus false negatives. 
As an additional measure, we lower $G$ of the CFAR detector to prevent false positives at fractional angles $\pm \alpha_{\textrm{max}}$ thanks to insufficient compression of objects. More concretely, objects $\boldsymbol{W}_{\pm \alpha_\textrm{max}} \boldsymbol{s}_O$ raise $\hat{\sigma}^2$ due to their spread larger than $2G$, which keeps the CFAR detector's computed energy {difference} below its detection threshold $\beta$. In Fig. \ref{fig:frfts}b we can observe how objects become more and more compressed as we approach $\alpha = \pm 90^\circ$; however, $\alpha = \pm 90^\circ$ is not included in the search space as $\alpha_\textrm{max} = 80^\circ$ in Fig. \ref{fig:frfts}b. 
A completely alternative approach would consist of unifying object and interference detection, treating detections at $\pm 90^\circ$ differently from detections at any other $\alpha$; we leave this idea for further research.

\subsection{Incomplete Interferences} \label{sec:incomplete}
\begin{figure}[!t]
\includegraphics[width=\columnwidth]{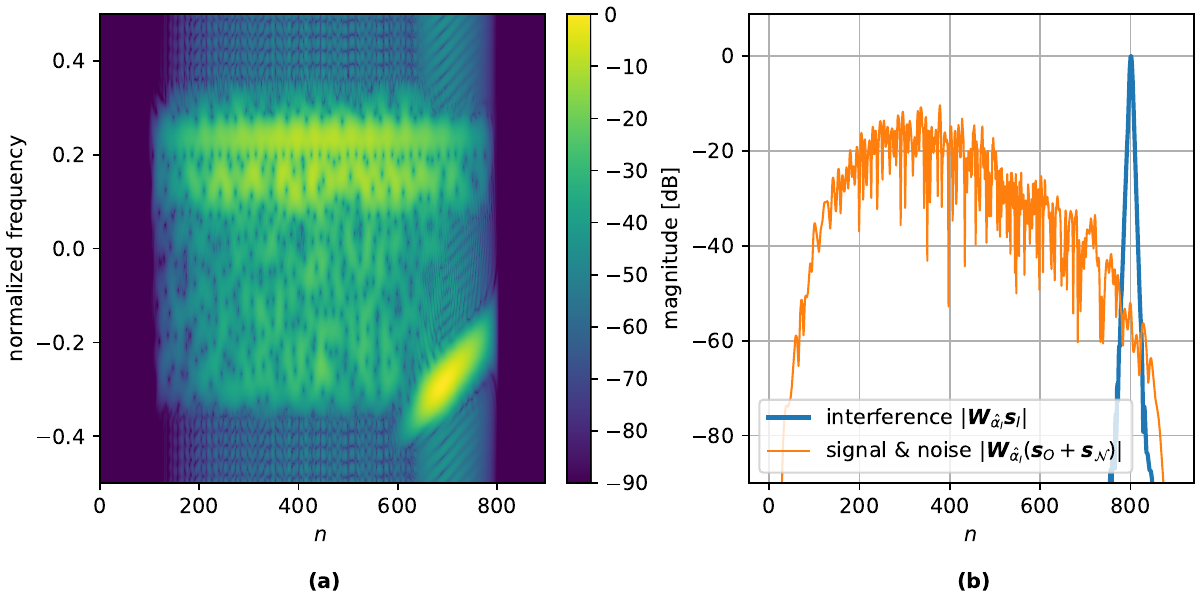}
\caption{\textbf{(a)} STFT of a radar signal with an incomplete interference \textbf{(b)} Magnitudes of the corresponding interference and ground truth signals after a DFrFT with $\hat{\alpha} \approx 45^\circ$. The signals have been normalized and padded with the technique described in Sec. \ref{sec:padding}.}
\label{fig:incomplete}
\end{figure}

Interferences have a certain starting and ending time, which are determined by the interferer and victim radars' parameters.  
One of the radar's fast-time sequence might end before the interference chirp crosses the entire receiver bandwidth,
resulting in an \textit{incomplete} interference. An interference also becomes incomplete if we zero another interference that crosses it in the time-frequency plane; {we discuss this scenario in Sec. \ref{sec:crossing}.}
The fractional representation $\boldsymbol{W}_{\hat{\alpha}_I}\boldsymbol{s}_I$ of such an interference does not contain all frequency components, i.e., it is not ideally compressed in the fractional domain. For very narrow interference bandwidths, our method fails to detect and therefore mitigate such interferences. Depending on the frequencies contained in $\boldsymbol{W}_{\hat{\alpha}_I}\boldsymbol{s}_I$, we can still detect incomplete interferences by increasing $G$ of the CFAR interference detector to account for their larger spread. However, as explained in Sec. \ref{sec:objs_vs_ints}, $G$ should be as small as possible to avoid misclassifying objects as interferences.
An example of an incomplete interference can be seen in Fig. \ref{fig:incomplete}.

{
\subsection{Crossing and non-crossing Interferences} \label{sec:crossing}
If the receive signal is corrupted by multiple interferences, we can distinguish between interference chirps that cross in the time-frequency domain and those that do not. Non-crossing interferences can be mitigated independently from one another, i.e., zeroing one interference chirp does not alter the appearance of the other. A prominent example for non-crossing interferences is the victim getting interfered by the same interferer multiple times within the same fast-time sequence, resulting in multiple interference chirps with the same chirp rate. If the interference chirps do cross in the time-frequency domain, then zeroing one interference will also zero a small portion of the other resulting in an incomplete interference, complicating its subsequent detection and mitigation. 
While it is possible to construct scenarios where zeroing one interference would render another interference undetectable, e.g., crossing interferences with almost identical chirp rates, we argue that crossing interferences are not problematic in the vast majority of scenarios. This unintended partial zeroing is analogous to zeroing parts of the object signal that overlap with interferences. An example for the quantitative impact of this side effect can be seen in Fig. \ref{fig:sig_ex}b by comparing the ground truth to the interference mitigated range-spectrum; in the interference mitigated range-spectrum, the objects are not as compressed as in the ground truth range-spectrum due to partial zeroing visible in Fig. \ref{fig:padding}c-d. However, this effect is so small that the objects can still be detected by a CFAR detector. Furthermore, partially zeroed interferences could be interpolated in future work using the methods described in Sec. \ref{sec:interpolation}. The runtime is identical for the crossing and the non-crossing case, namely $N_I + 1$ iterations of computing DFrFTs, CFAR-detection and zeroing.  
}

\section{Implementation and Computational Complexity} \label{sec:implementation}
As described in Sec. \ref{sec:mit}, our algorithm removes one interference at a time by computing a bank of DFrFTs, finding and classifying the global maximum followed by zeroing. The computational complexity is dominated by the bank of DFrFTs, which in turn depends on the DFrFT implementation used. We base our DFrFT implementation on the method introduced in \cite{vargas2005multiangle} which we call the MDFrFT; {they} showed that for a signal with length $N$, a bank of eigendecomposition-based centered DFrFTs with $N$ equally spaced fractional angles $\Bar{\boldsymbol{\alpha}}$ between $-180^\circ$ and $180^\circ$ has complexity $\mathcal{O}(N^2 \log N)$ instead of $\mathcal{O}(N^3)$. 
The MDFrFT $\bar{\boldsymbol{S}}$ of a signal $\boldsymbol{s}$ is computed as 
\begin{align} 
    \bar{\boldsymbol{S}}[p,n] & = \textrm{FFT}_p\{\bar{\boldsymbol{Z}}[p,n]\}, \label{eq:mafrft} \\ 
   \bar{\boldsymbol{Z}}[p,n] & = {\boldsymbol{V}^T[p,n]}(\boldsymbol{V}^T \boldsymbol{s})[p], \label{eq:pre_fft}
\end{align}
where $\textrm{FFT}_p\{\bar{\boldsymbol{Z}}[p,n]\}$ represents column-wise FFTs of $\bar{\boldsymbol{Z}}$ and $\boldsymbol{V}$ is the matrix of centered DFT eigenvectors. Each row $\bar{\boldsymbol{S}}[p]$ contains one of the DFrFTs with fractional angles $\boldsymbol{\Bar{\alpha}}$. 
Note that evaluating the FFTs in \eqref{eq:mafrft} has complexity $\mathcal{O}(N^2 \log N)$, while computing $\bar{\boldsymbol{Z}}$ has complexity $\mathcal{O}(N^2)$. The MDFrFT returns the exact same result as distinct eigendecomposition-based DFrFTs with angles $\bar{\boldsymbol{\alpha}}$. 

\subsection{EMDFrFT} \label{sec:MAFrFT}
{As discussed in Sec. \ref{sec:ang_res} and Sec. \ref{sec:n_angles}}, the number of evaluated fractional angles 
can be significantly lower than $N$ without impacting performance. 
However, the MDFrFT as proposed by \cite{vargas2005multiangle} always computes $N$ angles. 
{We generalize and reduce the computational complexity of the MDFrFT by computing $M$ equally spaced fractional angles $\boldsymbol{\alpha}_M$, where {$N \textrm{ mod } M = 0$}.}
This can easily be achieved by aliasing $\Bar{\boldsymbol{Z}}$ along its columns, which is equivalent to downsampling $\bar{\boldsymbol{S}}$ along its columns\cite{oppenheim1999discrete}, i.e., $\boldsymbol{S}[m, n] = \bar{\boldsymbol{S}}[mN/M, n]$, where
\begin{align}
    \boldsymbol{S}[m,n] & = \textrm{FFT}_m\{\boldsymbol{Z}[m,n]\}, \ \ m \in \{0,1,...,M-1\}, \label{eq:emdfrft} \\
    \boldsymbol{Z}[m,n] & = \sum_{l=0}^{\frac{N}{M}-1} \bar{\boldsymbol{Z}}[m + lM, n].  \label{eq:downsampling}
\end{align}
As we have now replaced $N$-point by $M$-point FFTs, the computational complexity of the required FFTs is $\mathcal{O}(NM \log M)$. For small $M$, the overall complexity of our EMDFrFT is $\mathcal{O}(N^2)$ since it is now dominated by the computation of $\bar{\boldsymbol{Z}}$. However, $\bar{\boldsymbol{Z}}$ can be calculated more efficiently by applying methods from \cite{vargas2005multiangle, de2019reduced, majorkowska2017low}, among others. Possible hardware architectures for such efficient eigendecomposition-based DFrFTs are described in \cite{bispo2024hardware}. Furthermore, \cite{erseghe2006efficient} {has} proposed an algorithm for computing the DFT {through} its eigendecomposition with complexity $\mathcal{O}(N \log N)$. If such an algorithm could be used to calculate \eqref{eq:pre_fft}, the MDFrFT and EMDFrFT would only consist of highly efficient divide-and-conquer based algorithms. In \cite{oswald2026onthe}, we halved the FFTs' computational burden within the MDFrFT by considering the symmetries of the DFT eigenvectors. Analog implementations of the DFrFT such as \cite{keshavarz2023real} are also promising directions for future research, as they completely circumvent the computational burden of digital DFrFT implementations. 

\subsection{Using the EMDFrFT for Interference Mitigation} \label{sec:emdfrft_int_mit}
The EMDFrFT computes DFrFTs for $M$ equally spaced fractional angles $\boldsymbol{\alpha}_M$ between $-180^\circ$ and $180^\circ$; {however}, as described in Sec. \ref{sec:mit}, we constrain our search space to {$N_{\boldsymbol{\alpha}}$} angles $\boldsymbol{\alpha}$ with $|\alpha| < \alpha_{\textrm{max}}$. We therefore retrieve the DFrFTs we include in the grid-search from the EMDFrFT as
\begin{equation}
    \boldsymbol{\alpha} = \{\alpha \in \boldsymbol{\alpha}_M, |\alpha| < \alpha_{\textrm{max}}\}. \label{eq:filter_angles}
\end{equation}
This means that 
\begin{equation}
    N_{\boldsymbol\alpha} = \Bigl\lfloor\frac{2\cdot M \cdot\alpha_{\textrm{max}}}{360^\circ}\Big\rfloor, \label{eq:n_alpha}
\end{equation}
with $\alpha_\textrm{max}$ in degrees.

If our algorithm performs multiple iterations, we can use the angle-additivity property of eigendecomposition-based DFrFTs to 
compute the next EMDFrFT directly on 
$\boldsymbol{d} \circ (\boldsymbol{W}_{\hat{\alpha}} \boldsymbol{s}$), where $\hat{\alpha} \in \boldsymbol{\alpha}$ is the found fractional angle in the previous iteration of our method. As $\boldsymbol{\alpha}_M$ consists of equally spaced fractional angles between $-180^\circ$ and $180^\circ$, the set of angles evaluated in the subsequent iteration is invariant to $\hat{\alpha}$ and therefore remains $\boldsymbol{\alpha}_M$. 
This means that we can implement \eqref{eq:filter_angles} by simply tracking the row indices of $\boldsymbol{S}$ that correspond to angles $\boldsymbol{\alpha}$ over consecutive iterations, which are cyclically row-wise shifted within $\boldsymbol{S}$ depending on $\hat{\alpha}$. More formally, in each iteration, we construct a binary matrix $\boldsymbol{M}$ such that 
{rows of $\boldsymbol{M} \odot \boldsymbol{S}$ corresponding to $\alpha \notin \boldsymbol{\alpha}$ are set to zero.}

If $M$ is a multiple of $4$, the DFT of $\boldsymbol{s}$ is computed as part of the EMDFrFT. Therefore, the range-FFTs in the radar signal processing chain can be removed and absorbed into the EMDFrFT, as is depicted in Fig. \ref{fig:proc_chain}. %
If our method performs multiple iterations, we additionally track the row index of the range-spectrum $m_{\textrm{RS}}$ within $\boldsymbol{S}$.

\subsection{Padding the Time-Fequency Representation} \label{sec:padding}

\begin{figure*}[!t]
\includegraphics[width=\textwidth]{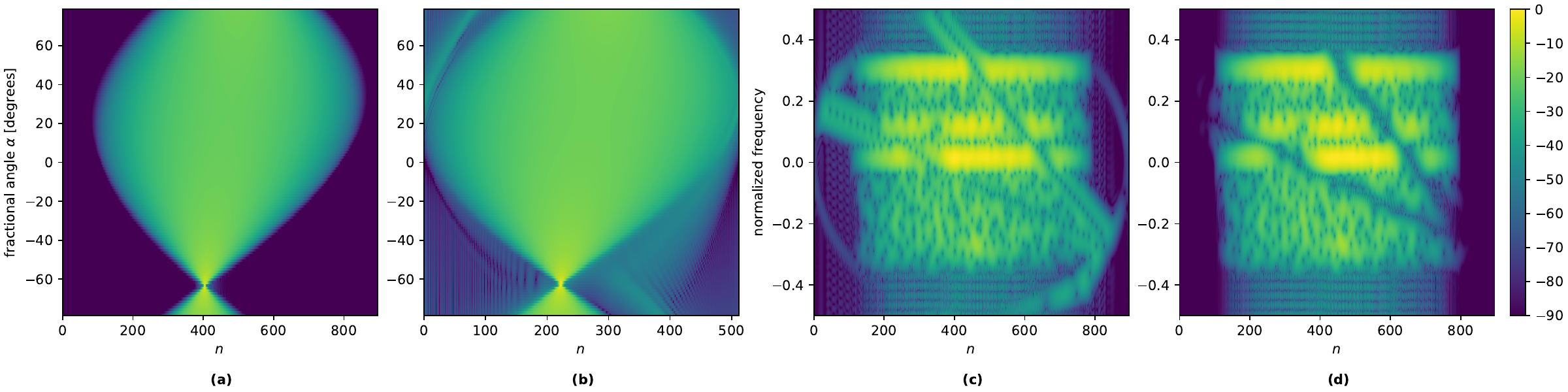}
\caption{Interference component from Fig. \ref{fig:sig_ex} with \textbf{(a)} and without \textbf{(b)} the padding introduced in Sec. \ref{sec:padding}. {STFTs} of the interference mitigated signal from Fig. \ref{fig:sig_ex} without {\textbf{(c)}} and with {\textbf{(d)}} a smoothening kernel; {\textbf{(c)} is the STFT of the interference mitigated signal in Fig. \ref{fig:sig_ex}a.} All plots are normalized and in dB.} 
\label{fig:padding}
\end{figure*}

In our implementation, we generate the DFT eigenvectors $\boldsymbol{V}$ as proposed in \cite{santhanam2008discrete}, which approximate the continuous FT eigenfunctions with concepts from quantum mechanics in finite dimensions. 
To deal with the approximation error, we found it helpful to zero-pad and oversample the input signals $\boldsymbol{s}$.
More concretely, we increase the sampling rate of the radar sensor's analog-digital converter to about $\gamma \cdot 2 f_c$, $\gamma = 1.32$, where $f_c$ is the cutoff frequency of the anti-aliasing filter. Furthermore, we prepend and append all processed radar signals $\boldsymbol{s}$ with $\lfloor\gamma \cdot N\rfloor$ zeroes after applying the windowing function. We derived $\gamma$ heuristically by fitting a circle around the original signal's time-frequency representation. 
Without padding, a DFrFT implementation using $\boldsymbol{V}$ from \cite{santhanam2008discrete} fails to properly transform signal components which are located in the corners of the signal's time-frequency representation. This results in suboptimal compression for LFM chirps that contain such components, as can be seen in Fig. \ref{fig:padding}b. These artefacts occur when a signal does not decay to zero at its boundaries for all fractional angles. 
The same LFM chirp with our proposed padding scheme is shown in Fig. \ref{fig:padding}a, which collapses to a single peak at approximately $-60^\circ$ as intended. We have also evaluated eigenvectors by \cite{candan2000discrete,clary2003shifted} and observed the same issues, which we mitigated with our padding scheme. 

Zeroing an interference induces broadband components into the signal's fractional representation $\boldsymbol{W}_{\hat{\alpha}} \boldsymbol{s}$, 
which might lead to artifacts after another subsequent DFrFT. Such artifacts with their circular-shaped appearance can be seen at the edges of Fig. \ref{fig:padding}c. 
If we perform padding and oversampling as introduced above, 
these artifacts are temporally and spectrally separated from the signal, and can therefore be removed after termination of our algorithm by cropping and low-pass filtering the interference mitigated range-spectrum. 
Without padding, the artifacts overlap with the object signal and cannot be removed anymore.
An alternative approach to deal with these artifacts consists of zeroing with a smoothening kernel, e.g., a raised cosine window, which depends on the location of the interference within the EMDFrFT. If we zero samples at fractional angles $\approx 45^\circ$ and close to the boundaries of the signal's fractional representation, we widen the smoothening kernel such that only low frequencies are suppressed. Therefore, the smoothening kernel prevents artifacts, and subsequent cropping and filtering can be avoided. An example is visualized in Fig. \ref{fig:padding}d. Our algorithm can be improved in future research by finding DFT eigenvectors that do not require such padding and oversampling. 

{
\subsection{Influence of Angular Resolution on Compression Strength} \label{sec:ang_res}
Let $s_I(t)$ be a continuous time-limited LFM chirp signal with chirp rate $k$, which is compressed most strongly when applying an FrFT with fractional angle $\hat{\alpha}_I = \text{arccot}(k)$\footnote{In this section, we restrict the range of all fractional angles to $[0, \pi)$ for concise notation.}. 
Furthermore, let $L$ be the width of $\mathcal{F}_{\tilde{\alpha}_I}s_I(t)$, with $\tilde{\alpha}_I = \hat{\alpha}_I + 90^\circ$, which transforms $s_I(t)$ into a time-limited sinusoid. Because an FrFT is, in simple terms, a rotation of the time-frequency plane, the width of $\mathcal{F}_{{\alpha}}s_I(t)$ is approximately $L \cdot |\sin(\alpha_\Delta)|$, where $\alpha_\Delta = |\hat{\alpha}_I - \alpha|$. Note that this approximation ignores the influence of side lobes which follow from $s_I(t)$ being time-limited; for instance, the actual width of $\mathcal{F}_{\hat{\alpha}_I}s_I(t)$ is $> 0$. If we now assume an idealized DFrFT $\boldsymbol{W}_{\alpha}$ which transforms a discrete LFM chirp signal $\boldsymbol{s}_I$, where $\boldsymbol{W}_{\tilde{\alpha}_I}\boldsymbol{s}_I$ spans the entire signal length $N$, we find that the width of that interference is approximately $\delta \coloneq \lceil N \cdot |\sin(\alpha_\Delta)|\rceil$ samples. 

Using these approximations we can now show, that for a given number of fractional angles $M$ we can afford to evaluate, the uniform angular resolution of the MDFrFT and EMDFrFT minimizes the worst-case interference width $\delta_{\text{max}}$ for all possible chirp rates $k$. For the MDFrFT and the EMDFrFT which evaluate fractional angles $\boldsymbol{\alpha}_M$, a mismatch $\alpha_\Delta$ is given by
\begin{equation*}
    \alpha_\Delta =|\hat{\alpha}_I - \alpha_{\text{closest}}| \leq \alpha_{\Delta,\max}, 
\end{equation*}
where $\alpha_{\text{closest}} \in \boldsymbol{\alpha}_M$ is the element minimizing the deviation from $\hat{\alpha}_I$. We have $\alpha_\Delta = \alpha_{\Delta,\max}$ if $\hat{\alpha}_I$ falls exactly in between the two closest elements of $\boldsymbol{\alpha}_M$, which means that $\alpha_{\Delta,\max} = 180^\circ/M$; however, this also means that, by the pigeonhole principle on continuous spaces, $\alpha_{\Delta,\max} > 180^\circ/M$ for any non-uniform spacing between elements of $\boldsymbol{\alpha}_M$ as there exists some interference angle where the deviation to $\alpha_{\text{closest}}$ is larger. Since $\delta_{\text{max}} =  \lceil N \cdot |\sin(\alpha_{\Delta,\max})|\rceil$ is growing monotonously for $0 <\alpha_{\Delta,\max} <\pi/2$, this uniform spacing also minimizes $\delta_{\text{max}}$. Therefore, the uniform spacing of $360^\circ /M$ provided by the MDFrFT and EMDFrFT minimizes $\delta_{\text{max}}$ for a given $M$ (and, transitively, $N_{\boldsymbol{\alpha}}$) such that all possible interferences are sufficiently compressed to be detectable. In Tab. \ref{tab:spreads} we can see the approximate worst-case widths as percentages of $L$, as well as $\delta_{\text{max}}$ for different $M$ and $N_{\alpha}$. 
\begin{figure}[!t]
\includegraphics[width=\columnwidth]{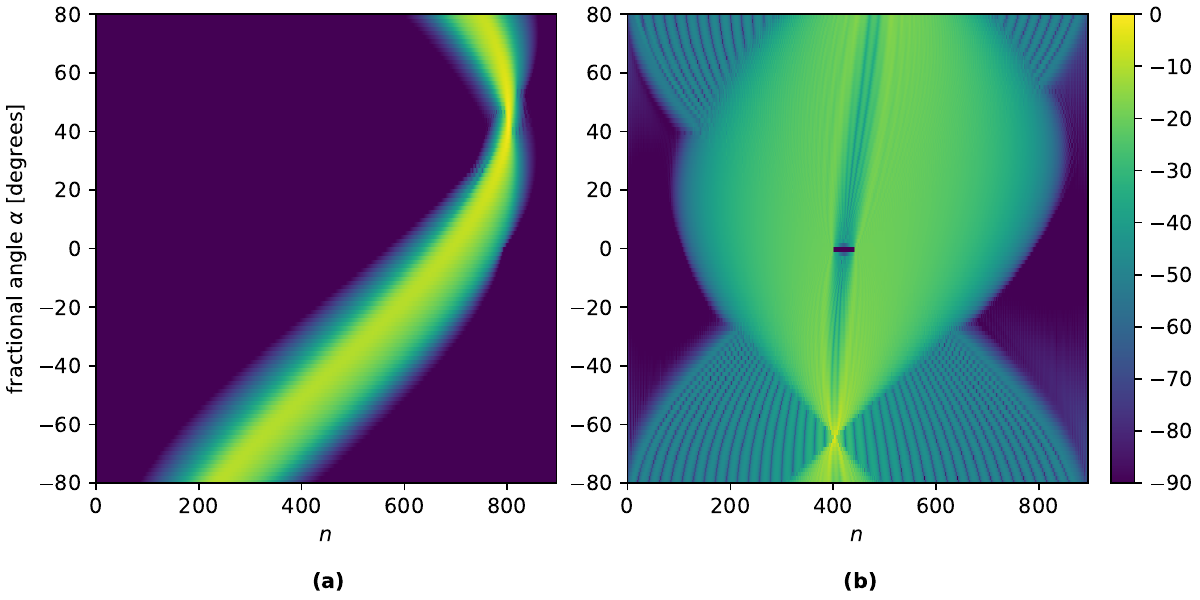}
\caption{{\textbf{(a)} EMDFrFT magnitudes of the interference component of Fig. \ref{fig:incomplete}. \textbf{(b)} EMDFrFT magnitudes of the signal in Fig. \ref{fig:padding}a after setting  some time-domain samples to zero, turning it into an incomplete interference. 
}}
\label{fig:incomplete_frfts}
\end{figure}

\begin{table}
    \centering
    {
    \caption{Approximate worst-case widths of interferences}
    \begin{tabular}{|c||c|c|c|c|}
    \hline
        M & \makecell{$N_{\boldsymbol{\alpha}}$ \\ if $\alpha_\textrm{max} = 80^\circ$}  & $\alpha_{\Delta,\max}$ & $| \sin(\alpha_{\Delta,\textrm{max}})|$ & \makecell{$\delta_{\text{max}}$ \\ if $N=512$} \\
        \hline
        \hline
        8 & 3 & $22.5^\circ$ & 38.27\% & 196  \\
        \hline
        16 & 7 & $11.25^\circ$ & 19.51\% & 100 \\
        \hline
        32 & 15 & $5.63^\circ$ & 9.80\% & 51 \\
        \hline
        64 & 29 & $2.81^\circ$ & 4.91\% & 26 \\
        \hline
        128 & 57 & $1.41^\circ$ & 2.45\% & 13   \\
        \hline
        256 & 113 & $0.70^\circ$ & 1.23\% & 7 \\
        \hline
        512 & 227 & $0.35^\circ$ & 0.61\% & 4 \\
        \hline
    \end{tabular}
    \label{tab:spreads}
    }
\end{table}

As already mentioned, these derivations only hold for idealized cases; in practical applications, the anti-aliasing filter and windowing are additional limiting factors for pulse-compressing interferences. If we extend our considerations to include the side lobes that follow from windowing or low-pass filtering chirps, we expect $\delta$ to be less sensitive to $\alpha_{\Delta}$ than in the idealized case, allowing us to lower $M$ with smaller effects on $\delta_{\max}$ than in Tab. \ref{tab:spreads}. More concretely, for a windowed and low-pass filtered chirp, the width of $\boldsymbol{W}_{\tilde{\alpha}_I}\boldsymbol{s}_I$ decreases while the width of $\boldsymbol{W}_{\hat{\alpha}_I}\boldsymbol{s}_I$ increases. The width of such an interference depending on $\alpha$ can be seen in Fig. \ref{fig:padding}a. The strength of this effect then depends on the low-pass filter's cut-off frequency as well as the timing and chirp rate of the interference. In extreme cases, such chirps might look like the one shown in Fig. \ref{fig:incomplete}; as visible in Fig. \ref{fig:incomplete_frfts}a, the width of these interferences is even less sensitive to $\alpha_\Delta$. This in turn means that a small value for $M$ is sufficient for treating such interferences, as none of the elements of $\boldsymbol{\alpha}_M$ can strongly compress the interference. 
In Fig. \ref{fig:incomplete_frfts}b we can see the EMDFrFT magnitudes of an incomplete interference which has partially been zeroed; this occurs if another interference, that had its maximum at that location, has previously been zeroed. Now, in addition to the peak, the interference is spread out across the fractional domain $\hat\alpha_I$ with low energy. In general, it is difficult to estimate the width of incomplete interferences.  
In Sec. \ref{sec:n_angles} we experimentally show that the EMDFrFT's angular resolution can be reduced to
$N_{\boldsymbol{\alpha}} = 29$ with only minor impacts on IM performance. For $N_{\boldsymbol{\alpha}} < 29$, choosing $N_{\boldsymbol{\alpha}}$ becomes a trade-off between computational complexity and performance. In Fig. \ref{fig:low_res}, we can see an example how reducing $N_{\boldsymbol{\alpha}}$ affects the compression strength of interferences. }

\subsection{Algorithm}
\begin{algorithm}[H]
\caption{Interference Mitigation using the EMDFrFT}
\begin{algorithmic}
\STATE 
\STATE {\textsc{IMfrac}}$(\boldsymbol{s})$: \COMMENT{$\boldsymbol{s}$ is a possibly interfered fast-time sequence}
\STATE $\text{initialize } \boldsymbol{M}$ \COMMENT{row indices of DFrFTs with angles $\boldsymbol{\alpha}$}
\STATE $\boldsymbol{s} \gets \boldsymbol{s} \odot \boldsymbol{w}$ \COMMENT{apply window function}
\STATE $\boldsymbol{s} \gets \mathrm{ZeroPad(\boldsymbol{s})}$ \COMMENT{optional, see Sec. \ref{sec:padding}}
\STATE $ \textbf{do} $
\STATE \hspace{0.5cm}$ \boldsymbol{S} \gets \textrm{EMDFrFT}(\boldsymbol{s})$  \COMMENT{see Sec. \ref{sec:MAFrFT}}
\STATE \hspace{0.5cm}$ \hat m, \hat n \gets \textrm{arg\,max}|\boldsymbol{M} \odot \boldsymbol{S}|$ %
\COMMENT{get indices of maximum}
\STATE \hspace{0.5cm}$ \boldsymbol{d} \gets \textrm{LO-CFAR}(\boldsymbol{S}[\hat m],\hat n)$ \COMMENT{$\boldsymbol{d}$ is a binary mask} %
\STATE \hspace{0.5cm}$ \boldsymbol{s} \gets \boldsymbol{d} \odot \boldsymbol{S}[\hat m]$ \COMMENT{set interfered samples to zero}
\STATE \hspace{0.5cm}$ \boldsymbol{M}, m_{RS} \gets \textrm{GetRows}(\hat m, \boldsymbol{d})$ \COMMENT{use angle additivity}
\STATE $ \textbf{while }\boldsymbol{d}\textrm{ contains a detection} $ \COMMENT{i.e., contains a zero}
\STATE $ \boldsymbol{S}[m_{RS}] \gets \textrm{Crop}(\boldsymbol{S}[m_{RS}]) $ \COMMENT{optional, see Sec. \ref{sec:padding}}
\STATE $ \boldsymbol{S}[m_{RS}] \gets \textrm{LowPass}(\boldsymbol{S}[m_{RS}]) $ \COMMENT{optional, see Sec. \ref{sec:padding}}
\STATE \textbf{return} $\boldsymbol{S}[m_{RS}]$ \COMMENT{interference mitigated range-spectrum}
\end{algorithmic}
\label{alg1}
\end{algorithm}

The final algorithm labeled \textit{IMfrac}\footnote{We provide Python code for our algorithm on \url{https://github.com/OsChri}.} that includes all optimizations is summarized in Alg. \ref{alg1}. All hyperparameters of our method are collected in Tab. \ref{tab:dfrft}.

{\section{Relationship to other Methods}\label{sec:properties}
In this section, we elaborate on how our proposed algorithm fits into the landscape of IM methods, which we have already briefly introduced in Sec. \ref{sec:intro} and Tab. \ref{tab:related}.}
\subsection{Relationship to Zeroing} \label{sec:zeroing}
Zeroing \cite{fischer2016untersuchungen} is one of the most common algorithms used for FMCW mutual IM due to its predictable and transparent behaviour. It works by detecting interferences in the time-domain input signal and simply setting all affected samples to zero. 
In the context of our proposed algorithm, zeroing can be viewed as a special case with 
$\boldsymbol{\alpha} = \{0^\circ\}$; 
therefore, we argue that our method is an improved version of zeroing.
The performance of zeroing highly depends on the interference detector. As the appearance of interferences in the time-domain is highly diverse, designing a robust interference detector is challenging; approaches include outlier or envelope change-point detection \cite{shimura2022advanced}, which assumes that interfered parts of the signal have higher amplitude compared to clean signal parts. By compressing the interference using the {EMDFrFT}, interference detection reduces to simple peak detection. In our method, we use the {EMDFrFT} as a means for both detecting and mitigating mutual interference; however, the {EMDFrFT} could also be used solely as an interference detector, which is then combined with some alternative treatment of the interference. We argue that, in fact, the {EMDFrFT} constitutes a highly performant LFM chirp interference detector due to its close relation to a bank of matched filters (see Sec. \ref{sec:matched}).
\begin{figure*}[!t]
\centering
\subfloat{\includegraphics[width=\textwidth]{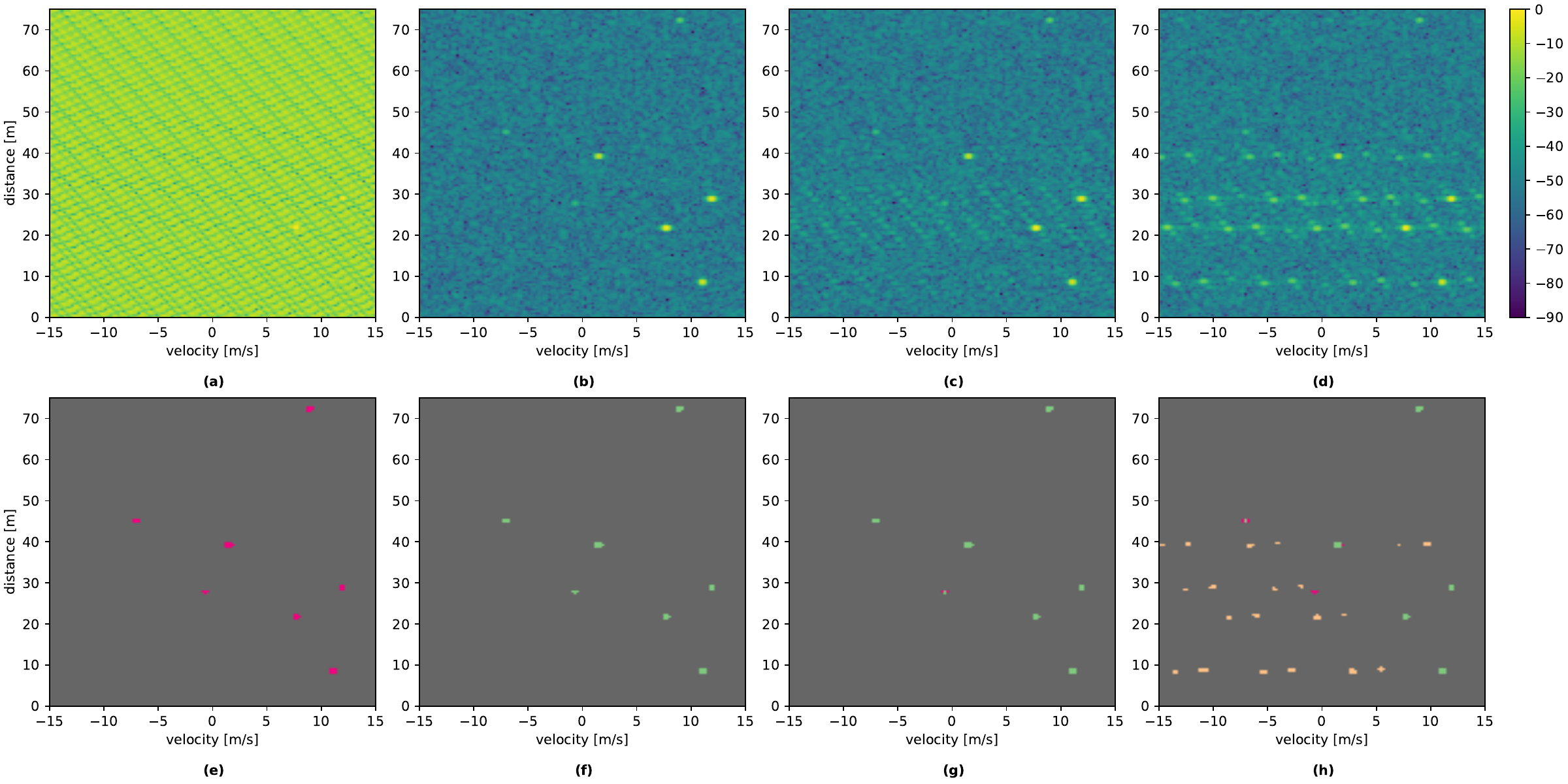}}
\caption{Examples of RD-maps. \textbf{(a)} interfered RD-map \textbf{(b)} corresponding ground truth RD-map \textbf{(c)} interference mitigated RD-map using our method \textbf{(d)} interference mitigated RD-map using zeroing with perfect interference detection. \textbf{(e-h)} object detection maps as retrieved by a CFAR object detector when applied to RD-maps (a-d). Gray, red, orange and green bins correspond to true negatives, false negatives, false positives and true positives, respectively.} 
\label{fig:rd}
\end{figure*}

Even with perfect interference detection, the performance of zeroing is still limited by the interference's chirp rate. Assuming an interference that crosses the victim radar's entire bandwidth, an interference with a lower chirp rate will affect a higher number of time-domain samples. Consequently, the number of zeroed time-domain samples increases, removing a larger proportion of the object signal as a side effect. When zeroing in the fractional domain, the loss of object signal components is much smaller and independent of the interference's chirp rate. This independence is especially relevant for subsequent object detections on RD-maps. The effects of IM on RD-maps are exemplified in Fig. \ref{fig:rd}. 
Zeroing typically leads to highly increased side lobes along the Doppler axis of a RD-map, as visible in Fig. \ref{fig:rd}d. This happens because the varying number of zeroed time-domain samples leads to object peaks having fluctuating amplitudes in the corresponding range-spectra. Computing the Doppler-FFT across these peaks 
then causes the aforementioned side lobes. IM based on the DFrFT does not suffer from this problem, strongly increasing object detection performance, as visible in and Fig. \ref{fig:rd}c and Fig. \ref{fig:rd}g. 

\subsection{Relationship to Matched Filtering} \label{sec:matched}
Matched filtering is a well established concept in signal processing and has a multitude of applications in fields like communications and radar. In pulse radar, matched filters are used to detect the presence of a transmit signal template in a noisy echo. In fact, LFM chirps are one of the most common signal templates in pulse radar, as their radar ambiguity function {has} desirable properties \cite{richards2005fundamentals}. Since LFM chirps are also observed as FMCW mutual interference, an approach for IM inspired by matched filtering is worth investigating. 

Our proposed method can be thought of as a bank of time-varying matched filter approximations, where each of the filters is tuned to an LFM chirp with a specific rate. While in theory one single filter would suffice to detect chirps with different rates, we use multiple filters in parallel and pick the one which compresses the interference the most to ensure reliable detection and precise mitigation. The filters are time-varying because DFrFT matrices are not Toeplitz matrices. 
Finally, matched filters are not necessarily invertible, which is a requirement for our approach based on multiple consecutive DFrFTs.
{
\subsection{Relationship to STFT-based Methods} \label{sec:relation_stft}
The STFT is one of the most popular transforms for processing nonstationary signals. Therefore, the STFT has also been used against FMCW radar mutual interference -- examples include \cite{muja2022interference, wang2021cfar, ristea2020fully}, among others. Conceptually, the STFT compresses an interference chirp into a line in its two-dimensional signal representation (e.g., see Fig. \ref{fig:incomplete}a); in contrast, like a matched filter, the DFrFT compresses a chirp into an impulse in its one-dimensional signal representation (e.g., see Fig. \ref{fig:incomplete}b), given that the DFrFT's fractional angle matches the chirp's rate. We argue that with the DFrFT one can detect and null LFM chirp interference more accurately than with the STFT, as the STFT’s time and frequency resolution is upper bounded by the Gabor uncertainty principle \cite{cohen1995}; by contrast, when computing a DFrFT, the transformed signal’s resolution remains the input signal length. Therefore, in opposition to Fig. \ref{fig:frfts}b, the STFTs {of the signals after IM in Fig. \ref{fig:padding}c-d cannot capture the precision of zeroing in the fractional Fourier domain.} However, the STFT may be used to treat any generic nonstationary interference, while the DFrFT, as used in our method, can only detect LFM chirp interference.

\subsection{Relationship to Method by Chen et al. \cite{chen2024interference}} \label{sec:relation_chen_et_al}
Independently from our work, \cite{chen2024interference} recently proposed a method for FMCW mutual IM, which also compresses, detects and zeroes interference chirps utilizing the FrFT. To iteratively compute approximate forward and inverse DFrFTs, the authors rely on the fast approximate FrFT \cite{ozaktas1996digital}, which has lower computational complexity than an exact eigendecomposition-based DFrFT, but does not offer angle-additivity, unitarity and reversibility. While unitarity could in principle be restored by renormalizing the transformed signal, the lack of reversibility accumulates additional errors in the processed radar signal during successive iterations of \cite{chen2024interference}. By contrast, we utilize the angle-additivity property of exact DFrFT implementations to skip the inverse DFrFT altogether. Moreover, using the multi-angle DFrFT \cite{vargas2005multiangle} allows us to compute exact DFrFTs more efficiently whilst absorbing the range-FFTs of the FMCW radar signal processing chain into our algorithm. Both of these optimizations reduce the computational complexity of our method. 
We argue that for the safety critical application of automotive radar, it is preferable to use exact DFrFT implementations, as the errors induced by approximate FrFT implementations reduce the trustworthiness of the IM algorithm’s output.

In \cite{chen2024interference}, the authors search for $\hat \alpha$ in an iterative manner using the golden-section search (GSS) algorithm; in contrast, our method implements a grid-search utilizing our efficient multi-angle DFrFT (EMDFrFT). As in \cite{chen2024interference} the signals are not renormalized after a fast approximate FrFT, the fractional angle found by the GSS in \cite{chen2024interference} is not guaranteed to be a maximum due the lack of unitarity, i.e., it does not necessarily compress an interference optimally. Furthermore, $\textrm{max}(|\boldsymbol{W}_\alpha \boldsymbol{s}|)$, as a function of $\alpha$, is not unimodal for any $\boldsymbol{s}$ except for pure LFM chirps, as $\boldsymbol{s}$ may contain objects, noise and multiple interferences. In that case, the GSS only finds a local maximum, which could be caused by weaker interferences, objects, noise or a superposition thereof. Meanwhile, our method does not make any assumptions on $\textrm{max}(|\boldsymbol{W}_\alpha \boldsymbol{s}|)$, and always returns its approximate global maximum; in Sec. \ref{sec:n_angles}, we show
that this approximation of the global maximum does not lead to performance degradation. Finally,
we argue that our grid search, which is computed in parallel, is more suitable than the iterative GSS
for real-time mutual IM.

Both \cite{chen2024interference} and our approach mitigate interferences in a loop until all interferences have been mitigated. However, \cite{chen2024interference} sets the maximum number of iterations a-priori based on the number of interferences in the input signal; our method does not have such a hyperparameter, as we solely rely on the LO-CFAR detector for terminating the algorithm. In \cite{chen2024interference}, they also terminate the algorithm when $\hat\alpha = \pi/2$. With this termination condition however, interferences with energies lower than the strongest object peak are not detected (assuming a unitary DFrFT implementation and that the golden-section search finds the global maximum).
By contrast, as described in Sec. \ref{sec:objs_vs_ints}, we detect interferences which are weaker than objects by tuning the hyperparameters $\alpha_\textrm{max}$ and the number of guard cells $G$ of the LO-CFAR.

After transforming $\boldsymbol{s}$ into the fractional domain with angle $\hat \alpha$, both methods use a CFAR detector to determine the interfered samples. Reference \cite{chen2024interference} uses a cell-averaging (CA) CFAR-detector on the entire transformed signal, while we use a LO-CFAR-detector on its maximum only. By classifying the global maximum only, we ensure that we detect and mitigate exactly one interference chirp per loop, and that this chirp is optimally compressed. Furthermore, we use a LO-CFAR-detector instead of a CA-CFAR detector to minimize the influence of other interferences on the CFAR-detector's noise floor estimate.

We compare \cite{chen2024interference} with our approach in terms of their performance in Sec. \ref{sec:experiments}.

\section{Experimental Setup} \label{sec:experimental_setup}

\subsection{Dataset} \label{sec:dataset}
We evaluate our method using a synthetic I/Q-modulated dataset consisting of 250 fast-time/slow-time sequences. The parameter values for the victim radar are collected in Tab. \ref{tab:ego}. {For every} fast-time/slow-time sequence, we generate interferences by uniformly sampling from the interferer parameter ranges summarized in Tab. \ref{tab:interferer}. Note that the number of interference chirps $N_I$ might be higher than the number of interferers, which is the case if the interferer and victim radars' frequency courses cross multiple times within the same victim radar fast-time sequence. We uniformly distribute between $0$ and $20$ objects across each RD-map with a maximum dynamic range of $60$ dB between objects. We retrieve the ground truth object detection maps by running a cell-averaging (CA) CFAR detector on the ground truth RD-maps, such that perfect reconstruction of the ground truth RD-maps results in a true positive rate and false negative rate of $1$ and $0$, respectively.
As our simulated dataset does not contain objects with negative ranges, we only consider positive ranges for all metrics described in Sec. \ref{sec:metrics}. An example for a RD-map from our dataset is shown in Fig. \ref{fig:rd}a-b. The dataset's distribution of SNRs and SINRs per RD-map can be seen in Fig. \ref{fig:ecdfs}b as \textit{ground truth} and \textit{no mitigation}, respectively.
\begin{table}[htbp]
\caption{Parameters of victim radar}
\begin{center}
\begin{tabular}{|c|c|}
\hline
\multicolumn{2}{|c|}{\textbf{Parameter}}  \\
\hline
Transmit starting frequency  & 79 \text{GHz} \\
\hline
Transmit bandwidth ($2 B_V$) & 0.25 \text{GHz} \\
\hline
Ramp duration ($ 2T_V$) & 12,8 $\mu s$ \\
\hline
Window type (range \& Doppler) & Hann \\
\hline
\# Fast-time samples ($N$) & 512 \\
\hline
\# Slow-time samples & 128 \\
\hline
\end{tabular}
\label{tab:ego}
\end{center}
\end{table}
\begin{table}[htbp]
\caption{Parameter ranges for interference signals}
\begin{center}
\begin{tabular}{|c||c|c|}
\hline
\textbf{Parameter} &\textbf{\textit{minimum}}& \textbf{\textit{maximum}} \\
\hline
\# Interferers  & 1 & 3 \\
\hline
Transmit starting frequency & 78.9 \text{GHz} & 79.0 \text{GHz} \\
\hline
Transmit bandwidth ($2 B_I$) & 0.2 \text{GHz} & 0.3 \text{GHz} \\
\hline
Ramp duration ($2 T_I$) & 10 $\mu s$ & 15 $\mu s$ \\
\hline
\# Slow-time samples & 100 & 156 \\
\hline
Dynamic range between interferers & 0 dB & 80 dB \\
\hline
\end{tabular}
\label{tab:interferer}
\end{center}
\end{table}

\vspace{-\baselineskip}
\subsection{{Evaluated} Methods}
\subsubsection{IMfrac}

We evaluate our proposed method with the parameters summarized in Tab. \ref{tab:dfrft}. 
Furthermore, we compare an oracle which has access to the isolated interference signals $\boldsymbol{s}_I$ as well as the ground truth clean signal $\boldsymbol{s}_O + \boldsymbol{s}_\mathcal{N}$. For each interference chirp $\boldsymbol{s}_I$, we find $\hat{\alpha}_I$ using the EMDFrFT. Then we set all samples $(\boldsymbol{W}_{\hat{\alpha}_I} \boldsymbol{s})[n]$ to zero where
\begin{equation}
   | \boldsymbol{W}_{\hat{\alpha}_I} \boldsymbol{s}_I|[n] > |\boldsymbol{W}_{\hat{\alpha}_I} (\boldsymbol{s}_O + \boldsymbol{s}_\mathcal{N})|[n].
\end{equation}
An example for such a comparison can be seen in Fig. \ref{fig:incomplete}b. The oracle establishes a performance upper bound for our algorithm that can be achieved if we estimate the object-plus-noise floor $\hat\sigma^2$ perfectly and choose $G$ optimally for each interference. 
We compare variants with and without the padding scheme introduced in Sec. \ref{sec:padding}. Padding the dataset described in Sec. \ref{sec:dataset} results in $N=896$. To make results comparable, we crop the padded output signals in the time and spectral domain such that $N=512$ for further processing. Note that we do not use a smoothening kernel for zeroing. For both variants, we set the CFAR detector's window size $\Phi$ such that $2\Phi$ covers the entire signal except for the cell under test and the guard cells. The CFAR detector's window is wrapped around the signal's edges. The parameters in Tab. \ref{tab:dfrft} have been set heuristically. $N_{\boldsymbol\alpha}$ is derived using \eqref{eq:n_alpha} from an EMDFrFT with $M=256$. We evaluate both variants of our method with $N_{\boldsymbol\alpha} = 113$ for better comparison, even though $M=256$ does not divide $N=896$ for the padded variant as required by the EMDFrFT. Therefore, when evaluating the padded variant, we replace the EMDFrFT with $113$ distinct DFrFTs.
\subsubsection{Method by Chen et al. \cite{chen2024interference}}
In Sec. \ref{sec:relation_chen_et_al} we analyzed the conceptual similarities as well as differences between our approach and \cite{chen2024interference}. 
We chose the same parameter values for the CA-CFAR interference detector within \cite{chen2024interference} as for our LO-CFAR peak classifier, which are listed in Tab. \ref{tab:dfrft}. We set the maximum number of iterations per fast-time sequence (labeled $M$ in \cite{chen2024interference}) to three to match the maximum number of interferers in our dataset. We also evaluate a variant of \cite{chen2024interference} which includes our padding scheme described in Sec. \ref{sec:padding}.}
\subsubsection{Zeroing}
We compare zeroing utilizing an envelope change-point interference detector. We also evaluate an oracle with perfect interference detection that zeroes all samples $\boldsymbol{s}[n]$ where $|\boldsymbol{s}_I|[n] > |\boldsymbol{s}_O + \boldsymbol{s}_\mathcal{N}|[n]$.
\subsubsection{Ramp Filtering}
We evaluate ramp filtering \cite{wagner2018threshold} applying a median-filter to the magnitudes of consecutive range-spectra. 
\begin{table}[htbp]
\caption{Parameters of IMfrac}
\begin{center}
\begin{tabular}{|c|c|}
\hline
\multicolumn{2}{|c|}{\textbf{Parameter}}  \\
\hline
\# Evaluated fractional angles ($N_{\boldsymbol{\alpha}}$) & 113 \\
\hline
Maximal fractional angle ($\alpha_{\textrm{max}}$) & $80^\circ$ \\
\hline
Window-size interference detector ($\Phi$) & $N/2 -G-1$ \\
\hline
\# Guard cells of interference detector ($G$) & 20 \\
\hline
Threshold of interference detector ($\beta$) & 20 dB \\
\hline
\end{tabular} \label{tab:dfrft}
\end{center} 
\end{table}
\subsection{Evaluation Metrics} \label{sec:metrics}
To gauge the performance of various IM methods, we evaluate their impact on object detections and reconstruction of interference-free RD-maps. 
\subsubsection{Mean-Squared Error (MSE)}
We use the MSE to compare the interference mitigated RD-map $\Tilde{\boldsymbol{S}}_{RD}$ to the ground truth  $\boldsymbol{S}_{RD}$. The MSE is defined as:
\begin{equation}
    \textrm{MSE} = \frac{1}{K} \sum_{K} |\Tilde{\boldsymbol{S}}_{RD} - \boldsymbol{S}_{RD}|^2, 
\end{equation}
where $K$ is the total number of RD-bins.

\subsubsection{Signal-to-interference-plus-noise Ratio (SINR)}
We compute the SINR as
\begin{equation}
    \textrm{SINR} = 10 \log\frac{\frac{1}{N_\mathcal{O}} \sum_{\{r,d\} \in \mathcal{O}} |\Tilde{\boldsymbol{S}}_{RD}[r,d]|^2}{\frac{1}{K-N_\mathcal{O}} \sum_{\{r,d\} \notin \mathcal{O}} |\Tilde{\boldsymbol{S}}_{RD}[r,d]|^2} , 
\end{equation}
where $\mathcal{O}$ is the set of all ground truth object bin locations $\{r,d\}$ containing $N_\mathcal{O}$ objects bins. $\mathcal{O}$ is aquired by running the CA-CFAR detector on $\boldsymbol{S}_{RD}$.
\subsubsection{Error Vector Magnitude}
The EVM describes the average proportion of the error vector to the ground truth object vector:
\begin{equation}
    \textrm{EVM} = \frac{1}{N_\mathcal{O}} \sum_{\{r,d\}\in \mathcal{O}} \frac{|\Tilde{\boldsymbol{S}}_{RD}[r,d] - \boldsymbol{S}_{RD}[r,d]|}{|\boldsymbol{S}_{RD}[r,d]|} 
\end{equation}
\subsubsection{False Alarm Rate (FAR)}
The false alarm rate of a predicted binary object detection map 
is computed as
\begin{equation}
    \textrm{FAR} = \frac{N_{FP}}{N_{FP} + N_{TN}}.
\end{equation}
The number of false positives $N_{FP}$ and true negatives $N_{TN}$ are acquired through a bin-wise comparison of the ground truth to the predicted object detection maps. 
\subsubsection{True Positive Rate}
In analogy to the FAR, the TPR is given by
\begin{equation}
    \textrm{TPR} = \frac{N_{TP}}{N_{TP} + N_{FN}},
\end{equation}
where $N_{TP}$ is the number of true positives and $N_{FN}$ the number of false negatives.
\subsubsection{F1-score}
The F1-score is a common metric to summarize the overall performance of a binary classifier. It is computed as
\begin{equation}
    \textrm{F1} = \frac{2N_{TP}}{2N_{TP} + N_{FP} + N_{FN}}.
\end{equation}
\section{Results} \label{sec:experiments}
\subsection{Performance} \label{sec:main_exp}
\begin{figure*}[!t]
\centering
\includegraphics[width=\textwidth]{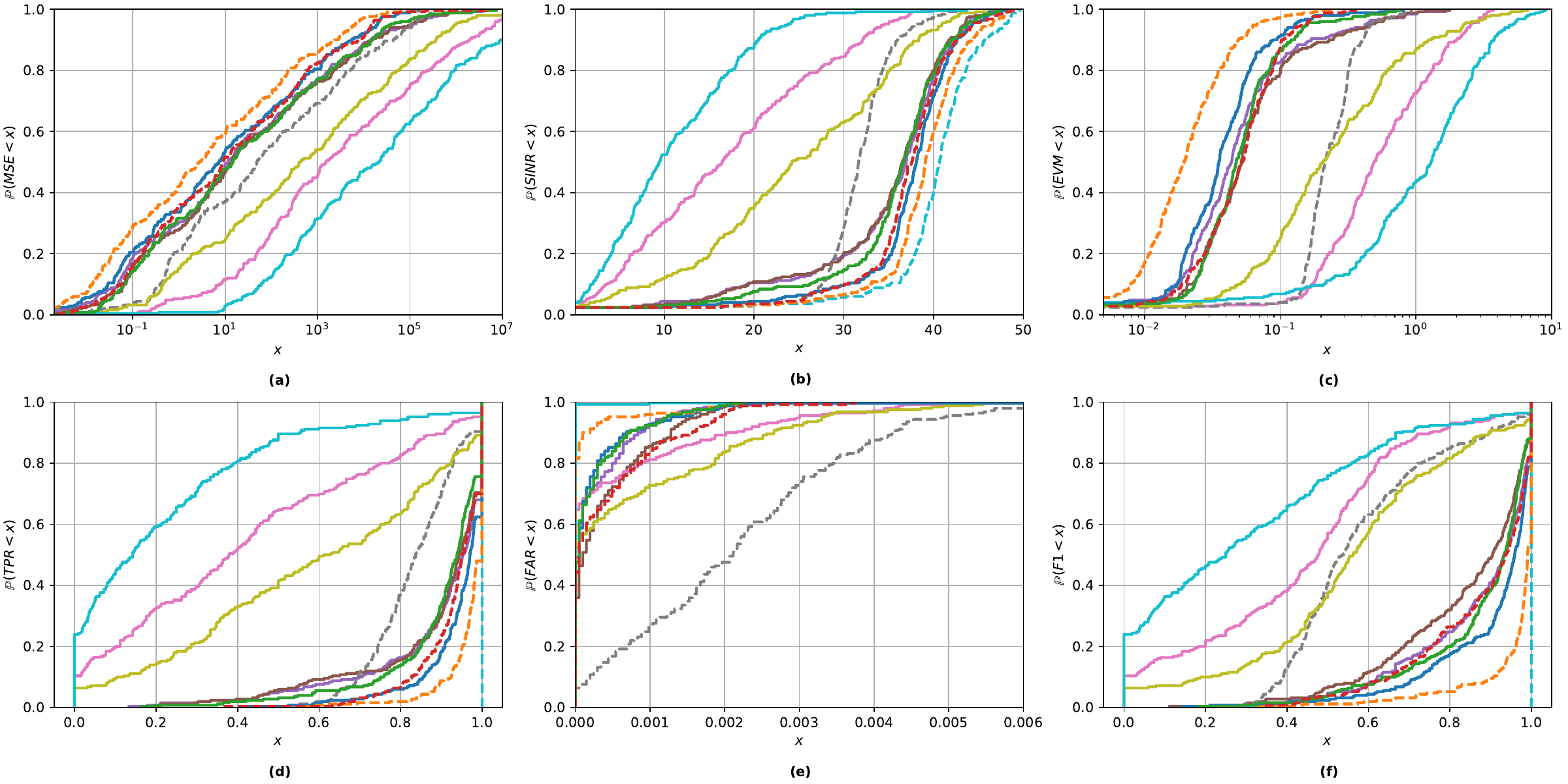}
\subfloat{\includegraphics[height=0.25in]{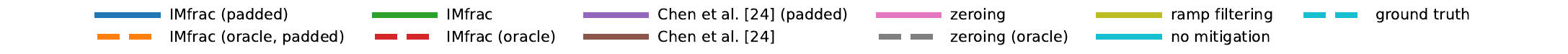}}%
\caption{Empirical cumulative density functions (ECDFs) of all evaluated metrics per range-Doppler map. The oracle methods are drawn with dashed lines. Note that we have zoomed into relevant parts of the ECDFs to better resolve close-by curves. {\textbf{(a)} MSE; \textbf{(b)} SINR; \textbf{(c)} EVM; \textbf{(d)} TPR; \textbf{(e)} FAR; \textbf{(f)} F1-score.}}  
\label{fig:ecdfs}
\end{figure*}

As we can see in Fig. \ref{fig:ecdfs}, our proposed \textit{IMfrac} performs best across all metrics. 
Padding generally improves the performance of our method, as interferences are detected more easily and artifacts can be removed after termination of our algorithm.
Performance improvements are larger for the oracle variants than for the CFAR-detection-based variants, as the oracle without padding sometimes falsely zeroes a high number of samples if an interference has not been properly compressed. On some metrics, the performances of the oracles are closely matched by the CFAR-based implementations. The method by Chen et al. \cite{chen2024interference} also outperforms zeroing and ramp filtering by a large margin, which highlights the overall potential of the FrFT for FMCW mutual IM. Furthermore, our padding scheme also improves the performance of \cite{chen2024interference}, as the approximation error of the fast approximate FrFT is smaller when the signal's energy is confined to the center of the time-frequency plane \cite{ozaktas1996digital}. We observe the typical effects of zeroing as explained in Sec. \ref{sec:zeroing}, where zeroing with oracle detection leads to a high TPR, but also a high FAR; {an} example can be seen in Fig. \ref{fig:rd}h. The performance gap between zeroing with oracle and envelope change-point detection highlights the difficulty of designing robust interference detectors in the time-domain. Ramp filtering outperforms zeroing with envelope change-point detection in our experiment. %

\subsection{Number of {F}ractional {A}ngles $N_{\boldsymbol{\alpha}}$} \label{sec:n_angles}

\begin{figure}[!t]
\includegraphics[width=\columnwidth]{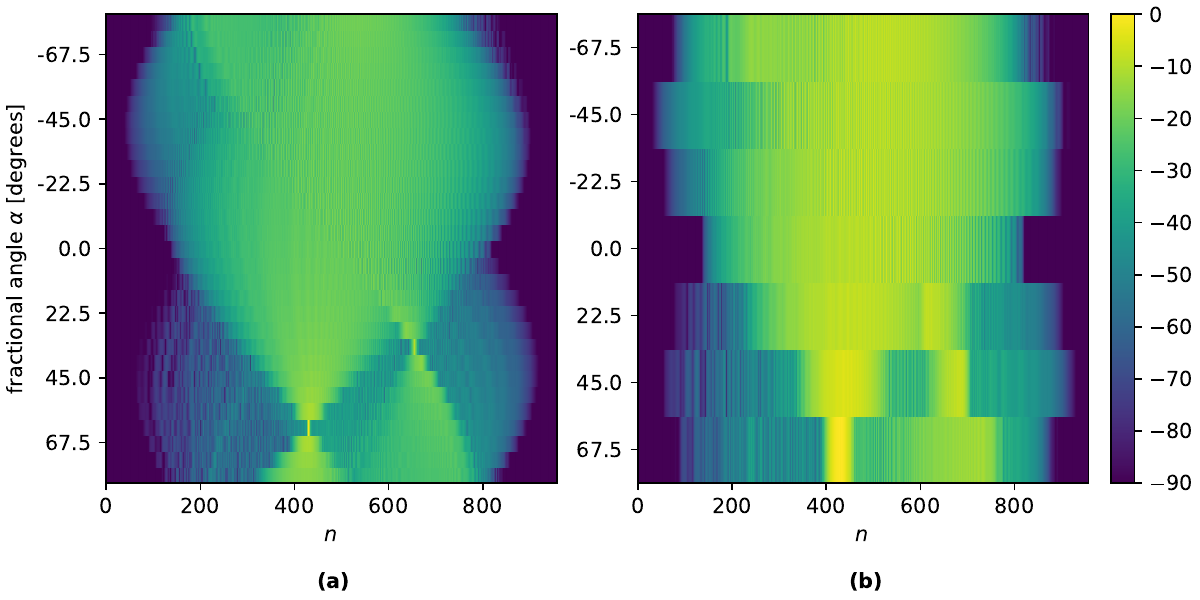}
\caption{{EMDFrFT magnitudes of the interfered signal in Fig. \ref{fig:sig_ex} with \textbf{(a)} $N_{\boldsymbol{\alpha}} = 29$ and \textbf{(b)} $N_{\boldsymbol{\alpha}} = 7$. Even though the angular resolution has been reduced compared to Fig. \ref{fig:frfts}a, where $N_{\boldsymbol{\alpha}} = 113$, the interferences are still sufficiently compressed in \textbf{(a)} so that they can be detected by the LO-CFAR detector. However, in \textbf{(b)}, the chirp rates of the interferences do not closely match any of the evaluated fractional angles, such that the interferences are not sufficiently compressed, leading to the performance degradation observed in Fig. \ref{fig:ecdf_angles}. As we are downsampling the angular resolution in \eqref{eq:downsampling} when computing an EMDFrFT, the fractional angles evaluated in \textbf{(b)} are a subset of the fractional angles evaluated in \textbf{(a)}, which in turn are a subset of the fractional angles evaluated in Fig. \ref{fig:frfts}a.
}}
\label{fig:low_res}
\end{figure}

\begin{figure*}[!t]
\includegraphics[width=\textwidth]{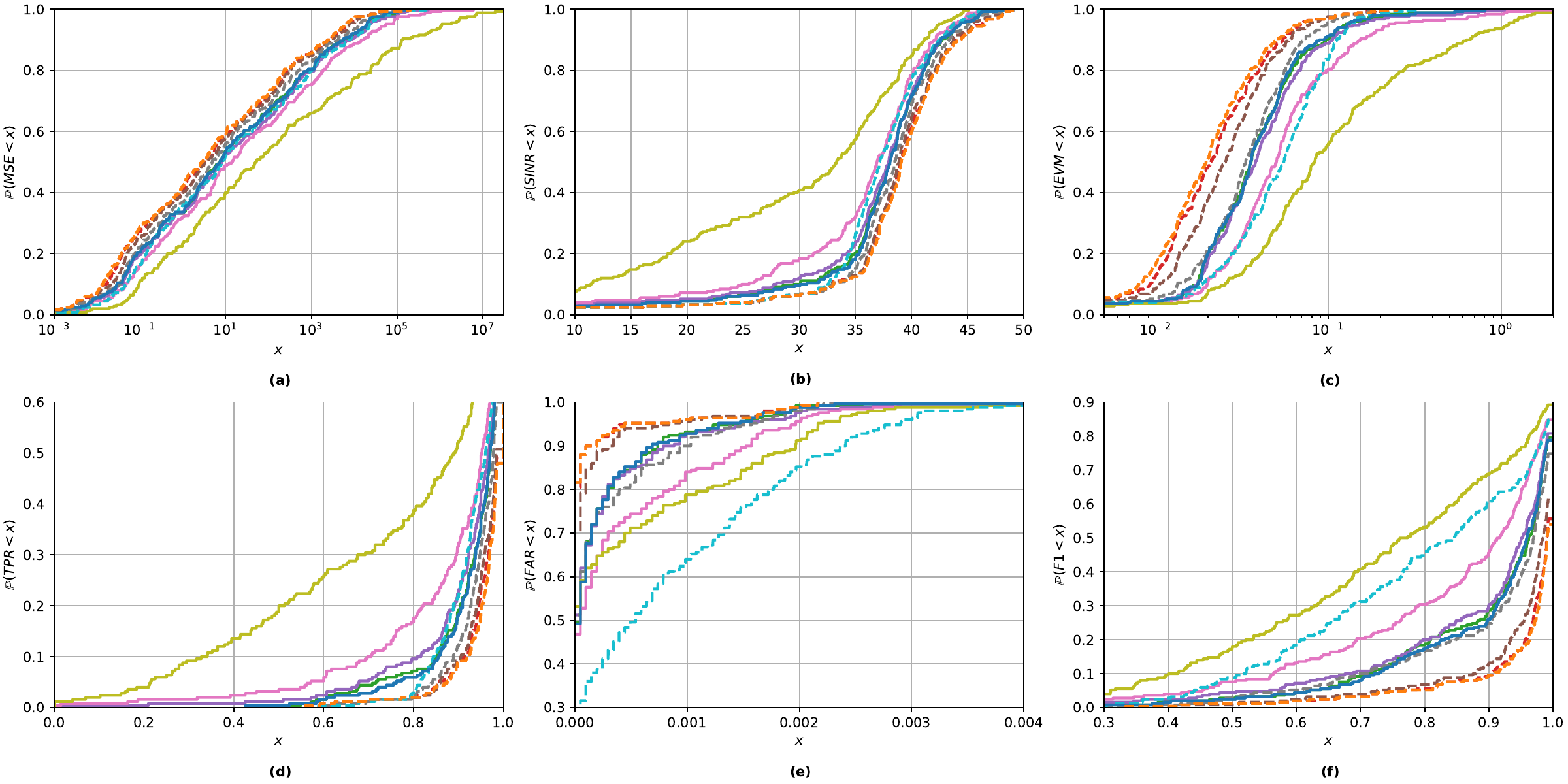} \\
\centering
\includegraphics[height=0.25in]{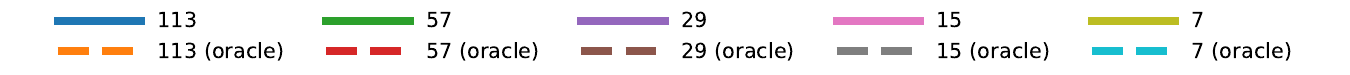}
\caption{Performance of our proposed method for different $N_{\boldsymbol{\alpha}}$. Note that we zoomed into relevant portions of the ECDFs to better resolve close-by curves. {\textbf{(a)} MSE; \textbf{(b)} SINR; \textbf{(c)} EVM; \textbf{(d)} TPR; \textbf{(e)} FAR; \textbf{(f)} F1-score.}} \label{fig:ecdf_angles} 
\end{figure*}

In this experiment, we evaluate our method for different $N_{\boldsymbol{\alpha}}$. We set $M = \{256, 128,64,32,16\}$ and derive the corresponding $N_{\boldsymbol{\alpha}} = \{113,57,29,15,7\}$ using \eqref{eq:n_alpha}, while keeping all other parameters the same as in Tab. \ref{tab:dfrft}. We apply the padding scheme from Sec. \ref{sec:padding} without a smoothening kernel. The results for $N_{\boldsymbol{\alpha}} = 113$ are the same as in Sec. \ref{sec:main_exp} and can be seen in Fig. \ref{fig:ecdf_angles}. As expected, the performance of the oracles is steadily worsening with decreasing $N_{\boldsymbol{\alpha}}$; however, when a CFAR-detector is used for interference detection, performance remains roughly constant for $N_{\boldsymbol{\alpha}} = \{113,57,29\}$. Therefore, for $N = 896$, $M$ can be reduced from $896$ to $64$ while only marginally reducing performance. 
This corresponds to a reduction of the number of operations for the FFTs in \eqref{eq:emdfrft} from roughly $7.87 \cdot 10^6$ for the MDFrFT \cite{vargas2005multiangle} to $3.44 \cdot 10^5$ for the EMDFrFT, i.e., by a factor of approximately $23$. 
For $N_{\boldsymbol{\alpha}} = \{15,7\}$, the performance gap between the oracle and the CFAR interference detector steadily widens; {this} gap could be narrowed by developing a more elaborate interference detector in future research, so that $M$ can be reduced even further. For decreasing $N_{\boldsymbol{\alpha}}$ our method becomes more and more similar to zeroing, which explains its increasingly high false alarm rate. {An interfered signal's EMDFrFT for different $N_{\boldsymbol\alpha}$ can be seen in Fig. \ref{fig:low_res}.}

\section{Conclusion {and Future Work} \label{sec:conclusion}} 
In this work we have presented a novel method for FMCW radar mutual interference mitigation which is based on the discrete fractional Fourier transform (DFrFT). Our method performs multiple consecutive DFrFTs to detect and null interferences in the fractional domain. We have analyzed the properties of our method, and also provided an implementation {optimized for real-time, resource-constrained and safety critical applications} which makes use of our new efficient multi-angle discrete fractional Fourier transform (EMDFrFT) and the angle-additivity property of the DFrFT. We also proposed a practical method to improve the chirp compression capabilities of the eigendecomposition-based DFrFT by \cite{santhanam2008discrete, candan2000discrete, clary2003shifted}, among others. All of these contributions lead to {a} simple algorithm which achieves competitive performance across all our considered metrics on an I/Q-modulated dataset. Throughout this paper, we have indicated potential future improvements for our method. In future work, we plan comparisons to data-driven approaches such as \cite{rock2020deep, fuchs2021complex}.
We plan to extend our algorithm to real-valued radar data, where the interferences appear as real-valued LFM chirps. Even though real-valued chirps are not basis functions of the FrFT, the exact same Algorithm \ref{alg1} can in principle be applied to real-valued radar data.
{\subsection{Extension to More Realistic Signal Models}
In Sec. \ref{sec:sig_model}, we have introduced a signal model which neglects multipath components as well as imperfections of real radar systems. More realistic signal models could be evaluated in future research; 
in this section, we briefly describe how we expect our method to generalize to real-world environments and radar sensors. 

Our model could be extended to multipath environments, where instead of an impulse we find the channel’s impulse response in an interference’s optimal fractional domain.
    
Our algorithm could also be used with non-ideal anti-aliasing filters, where the aliased components of the interference chirp may be included in the kernel of the CFAR interference detector. 
    
Another common imperfection of radar systems is phase noise; we expect that phase noise in an interference chirp leads to a widening of the compressed interference after a DFrFT with the optimal fractional angle, which is akin to the spectral regrowth of sinusoids. As long as this widening of the interference peak is not excessive, we conjecture that our method can be applied to interferences corrupted by phase noise.
    
The dynamic range of real-world radar systems is finite, and strong interferences might saturate the receiver. {If we model the receiver as a static nonlinearity which clips at some given input level, an interference saturating the receiver would induce intermodulation products as well as spectral regrowth \cite{cripps2006rf}. The intermodulation products appear as additional LFM chirps with chirp rates depending on the original interferences; therefore, we argue that our method could also mitigate these intermodulation products. Spectral regrowth, on the other hand, widens the spectrum of the interference chirps, hindering optimal compression with the DFrFT. More research is necessary to determine acceptable levels of saturation.} 
An additional strategy could be to replace the LO-CFAR interference detector with a more elaborate detector that operates on the entire bank of DFrFTs $\boldsymbol{S}$, as the different peaks in $|\boldsymbol{S}|$ corresponding to the different harmonics always appear in predictable patterns.

\subsection{Interpolating the Zeroed Signal Components} \label{sec:interpolation}
In our method, we set the compressed interferences to zero; however, as described in Sec. \ref{sec:mit}, we also unintentionally remove object signal components in the process. Meanwhile, publications such as \cite{marvasti2012sparse, neemat2018interference, wang2021matrix}, among others, propose schemes to interpolate zeroed components using estimates of the object signal; more concretely, they are used as a post-processing step, typically after time-domain zeroing using some interference detector \cite{fischer2016untersuchungen}. 
 Our method could be improved in future research by combining it with such an interpolation scheme. In the case of time-domain interpolation algorithms however, we would first need to adapt these algorithms to zeroing the fractional Fourier domain.
 For instance, a well-known interpolation scheme is the iterative method with adaptive thresholding (IMAT) \cite{marvasti2012sparse}, where the previously zeroed time-domain samples are interpolated using object peak detections in the range-spectrum. 
 While IMAT, in its original formulation, iteratively computes inverse and forward DFTs interpolating the signal gaps as object peaks are detected, we can adapt this procedure to our method by simply replacing the DFTs with DFrFTs; we would therefore iterate between the range-spectrum and the fractional domain where zeroing was performed, and generate an interpolation signal which consists of a slice of LFM chirps instead of sinusoids. This approach could even be extended to interpolate interferences that have previously been partly zeroed, such that they are not incomplete anymore. A weakness of IMAT and related methods is that the interpolation's quality depends on the number of previously zeroed time-domain samples, as object signal estimates are formed from the remainder of the radar signal; however, this portion of the signal is short if the interference is long in the time-domain, e.g., if its chirp rate is low.
 {As discussed in Sec. \ref{sec:zeroing}, the number of zeroed samples is lower and also practically independent of the interference's chirp rate when using our method.} Therefore, we expect this adapted version of IMAT to have better and more constant performance for all possible interference chirp rates compared to its original formulation.

\bibliographystyle{IEEEtran}
\bibliography{references}

\end{document}